\documentclass[onecolumn, 12pt]{IEEEtran}
\usepackage{algorithm}
\usepackage[displaymath,mathlines]{lineno}
\usepackage{algorithmic}
\usepackage{cite}
\usepackage{epsfig}
\usepackage{graphics}
\usepackage{amsmath}
\usepackage{amssymb}
\usepackage{latexsym}
\usepackage[]{fontenc}
\usepackage{psfrag}
\usepackage{subfigure}
\usepackage[]{fontenc}

\DeclareGraphicsExtensions{ps,eps,ps.gz}

\def\bh{{\boldsymbol{h}}}

\newcommand{\transp}{^{\rm T}}

\newcommand{\p}[1]{\mathop{\mbox{\it p} } }

\renewcommand{\vec}[1]{\ensuremath{\boldsymbol{#1}}}
\newcommand{\be}{\begin{equation}}
\newcommand{\ee}{\end{equation}}
\newcommand{\ba}{\begin{array}}
\newcommand{\ea}{\end{array}}
\newcommand{\bea}{\begin{eqnarray}}
\newcommand{\eea}{\end{eqnarray}}
\newcommand{\bean}{\begin{eqnarray*}}
\newcommand{\eean}{\end{eqnarray*}}

\newcommand{\fw}{_{\mathrm{f}}}
\newcommand{\pil}{_{\mathrm{p}}}
\newcommand{\fwk}{_{\mathrm{f}k}}

\newcommand{\HT}{^{\rm H}}

\renewcommand{\Re}{\mathop{\rm Re}}

\long\def\symbolfootnote[#1]#2{\begingroup%
\def\thefootnote{\fnsymbol{footnote}}\footnote[#1]{#2}\endgroup}

\begin{document}


\title{Scaling up MIMO:  Opportunities and Challenges with Very Large Arrays}

\author{ \authorblockN{Fredrik Rusek{$^\dagger$}\footnote{$^\dagger$
      Dept. of Electrical and Information Technology, Lund University,
      Lund, Sweden}, Daniel Persson{$^\ddagger$}\footnote{$^\ddagger$
      Dept. of Electrical Engineering (ISY), Link\"{o}ping University,
      Sweden}, Buon Kiong Lau{$^\dagger$}, Erik
    G. Larsson{$^\ddagger$}, Thomas
    L. Marzetta{$^{\S}$}\footnote{$^{\S}$ Bell Laboratories,
      Alcatel-Lucent, Murray Hill, NJ}, Ove Edfors{$^\dagger$}, and Fredrik
    Tufvesson{$^\dagger$}\footnote{Contact authors:
      Fredrik Rusek \emph{fredrik.rusek@eit.lth.se} and Daniel Persson
      \emph{daniel.persson@isy.liu.se}}} }

\maketitle



\vspace*{-15mm}
\section{Introduction}
MIMO technology is becoming mature, and incorporated into emerging
wireless broadband standards like LTE \cite{LTE}. For example, the LTE
standard allows for up to 8 antenna ports at the base
station. Basically, the more antennas the transmitter/receiver is
equipped with, and the more degrees of freedom that the propagation
channel can provide, the better the performance in terms of data rate
or link reliability. More precisely, on a quasi-static channel where a
codeword spans across only one time and frequency coherence interval,
the reliability of a point-to-point MIMO link scales according to
$\mbox{Prob(link outage)}\sim \mbox{SNR}^{-n_tn_r}$ where $n_t$ and
$n_r$ are the numbers of transmit and receive antennas, respectively,
and SNR is the Signal-to-Noise Ratio.  On a channel that varies
rapidly as a function of time and frequency, and where circumstances
permit coding across many channel coherence intervals, the achievable
rate scales as $\min(n_t, n_r)\log(1+\mbox{SNR})$.  The gains in multiuser
systems are even more impressive, because such systems offer the
possibility to transmit simultaneously to several users and the
flexibility to select what users to schedule for reception at any
given point in time \cite{tv:05}.

The price to pay for MIMO is increased complexity of the hardware
(number of RF chains) and the complexity and energy consumption of the
signal processing at both ends. For point-to-point links, complexity
at the receiver is usually a greater concern than complexity at the
transmitter. For example, the complexity of optimal signal detection
alone grows exponentially with $n_t$ \cite{Larsson2009,jalden}. In
multiuser systems, complexity at the transmitter is also a concern
since advanced coding schemes must often be used to transmit
information simultaneously to more than one user while maintaining a
controlled level of inter-user interference. Of course, another cost of MIMO is
that of the physical space needed to accommodate the antennas,
including rents of real estate.

With \emph{very large MIMO}, we think of systems that use antenna
arrays with an order of magnitude more elements than in systems being
built today, say a hundred antennas or more.  Very large MIMO entails an unprecedented number of antennas simultaneously serving a much smaller number of terminals. The disparity in number emerges as a desirable operating condition and a practical one as well. The number of terminals that can be simultaneously served is limited, not by the number of antennas, but rather by our inability to acquire channel-state information for an unlimited number of terminals. Larger numbers of terminals can always be accommodated by combining very large MIMO technology with conventional time- and frequency-division multiplexing via OFDM. Very large MIMO arrays is a new
research field both in communication theory, propagation, and
electronics and represents a paradigm shift in the way of thinking
both with regards to theory, systems and implementation.  The ultimate
vision of very large MIMO systems is that the antenna array would
consist of small active antenna units, plugged into an (optical)
fieldbus.

We foresee that in very large MIMO systems, each antenna unit uses
extremely low power, in the order of mW.  At the very minimum, of
course, we want to keep total transmitted power constant as we
increase $n_t$, i.e., the power per antenna should be $\propto
1/n_t$. But in addition we should also be able to back off on the
\emph{total} transmitted power. For example, if our antenna array were
serving a single terminal then it can be shown that the total power can be
made inversely proportional to $n_t$, in which case the power required
per antenna would be $\propto 1/n_t^2$.  Of course, several
complications will undoubtedly prevent us from fully realizing such
optimistic power savings in practice: the need for multi-user
multiplexing gains, errors in Channel State Information (CSI), and
interference.  Even so, the prospect of saving an order of
magnitude in transmit power is important because one can achieve
better system performance under the same regulatory power constraints.
Also, it is important because the energy consumption of cellular base
stations is a growing concern. As a bonus, several expensive and bulky
items, such as large coaxial cables, can be eliminated
altogether. (The  coaxial cables used for tower-mounted base stations today
are up to four centimeters in diameter!) Moreover, very-large MIMO
designs can be made extremely robust in that the failure of one or a
few of the antenna units would not appreciably affect the
system. Malfunctioning individual antennas may be hotswapped.  The
contrast to classical array designs, which use few antennas fed  
from a high-power amplifier, is significant.

So far, the large-number-of-antennas regime, when $n_t$ and $n_r$ grow
without bound, has mostly been of pure academic interest, in that some
asymptotic capacity scaling laws are known for ideal situations. More
recently, however, this view is changing, and a number of practically
important system aspects in the large-$(n_t,n_r)$ regime have been
discovered. For example, \cite{Marzetta10} showed that asymptotically
as $n_t\to\infty$ and under realistic assumptions on the propagation
channel with a bandwidth of 20 MHz, a
time-division multiplexing cellular system may accommodate more than
40 single-antenna users that are offered  a net \emph{average} throughput
of 17 Mbits per second both in the reverse (uplink) and the forward
(downlink) links, 
and a throughput of 3.6 Mbits per second \emph{with 95\% probability}!
These rates are achievable \emph{without
  cooperation among the base stations} and by relatively rudimentary
techniques for CSI acquisition based on uplink pilot measurements.

Several things happen when MIMO arrays are made large. First, the
asymptotics of random matrix theory kick in.  This has several consequences. Things that were
random before, now start to look deterministic.  For example, the
distribution of the singular values of the channel matrix approaches a
deterministic function
\cite{randommtrx}.  Another fact is that very tall or very wide matrices tend to be very well conditioned.  Also when dimensions are large, some matrix operations 
such as inversions can be done fast,
by using series expansion techniques (see the sidebar).  
In the limit of an infinite number of antennas at the base station, but with a single antenna per user, then linear processing in the form
of maximum-ratio combining for the uplink (i.e., matched filtering with the channel vector, say $\bh$) and  maximum-ratio transmission (beamforming with $\bh\HT/||\bh||$) on  the downlink is optimal.  This 
resulting processing  is
reminiscent of time-reversal, a technique used for focusing electromagnetic or 
acoustic waves \cite{LRT2004,TWF1996}.

The second effect of scaling up the dimensions is that thermal noise can be
averaged out so that the system is predominantly limited by
interference from other transmitters. This is intuitively clear for the uplink, since coherent averaging offered
by a  receive antenna array eliminates quantities that are uncorrelated between the antenna elements, that is, thermal noise in particular.  This effect is less obvious on the downlink, however.  
Under certain circumstances,
the performance of a very large array becomes limited by interference arising from re-use
of pilots in neighboring cells. In addition, choosing pilots in a
smart way does not substantially help as long as the coherence time of
the channel is finite.  In a Time-Division Duplex (TDD) setting, this effect was quantified in
\cite{Marzetta10}, under the assumption that the channel is
reciprocal  and that the base stations estimate the downlink channels
by using uplink received pilots.  

Finally, when the aperture of the array grows, the resolution of the array increases.  This means that one can resolve individual scattering centers with unprecedented precision. Interestingly, as we will see later on, the communication performance of the array in the large-number-of-antennas regime depends less on the actual statistics of the propagation channel but only on the aggregated properties of the propagation such as asymptotic orthogonality between channel vectors associated with distinct terminals. 

Of course, the number of antennas in a practical system cannot be arbitrarily large owing to physical constraints.  Eventually, when letting $n_r$ or $n_t$ tend to infinity, our mathematical models for the physical reality will break down.  For example, the aggregated received power would at some point exceed the transmitted power, which makes no physical sense.  But long before the mathematical models for the physics break down, there will be substantial engineering difficulties.  So, how large is ``infinity'' in this paper?   The answer depends on the precise circumstances of course, but in general, the asymptotic results of random matrix theory are accurate even for relatively small dimensions (even 10 or so). In general, we think of systems with at least a hundred antennas at the base station, but probably less than a thousand. 

Taken together, the arguments presented motivate entirely new
theoretical research on signal processing and coding and network
design for very large MIMO systems. This article will survey some of these
challenges. In particular, we will discuss ultimate
information-theoretic performance limits, some practical algorithms,
influence of channel properties on the system, and practical
constraints on the antenna arrangements.

\subsection{Outline and key results}

The rest of the paper is organized as follows.  We start with a brief
treatment of very large MIMO from an information-theoretic
perspective. This provides an understanding for the fundamental limits
of MIMO when the number of antennas grows without bound. Moreover, it
gives insight into what the optimal transmit and receive strategies
look like with an infinite number of antennas at the base station.  It
also sets the stage for the ensuing discussions on realistic
transmitter and receiver schemes.

Next, we look at antennas and propagation aspects of large MIMO. First
we demonstrate how and why maximum-ratio transmission beamforming can
focus power not only in a specific \emph{direction} but to a given
\emph{point} in space and we explain the connection between this
processing and time-reversal. We then discuss in some detail mutual
coupling and correlation and their effects on the channel capacity,
with focus on the case of a large number of antennas. In addition, we
provide results based on measured channels with up to $128$ antennas.

The last section of the paper is dedicated to transmit and receive
schemes. Since the complexity of optimal algorithms scales with the
number of antennas in an unfavorable way, we are particularly
interested in the structure and performance of approximate,
low-complexity schemes. This includes variants of linear processing
(maximum-ratio transmission/combining, zero-forcing, MMSE) and
algorithms that perform local searches in a neighborhood around
solutions provided by linear algorithms. In this section, we also
study the phenomenon of \emph{pilot contamination}, which occurs when
uplink channel estimates are corrupted by mobiles in distant cells
that reuse the same pilot sequences. We explain when and why pilot
contamination constitutes an ultimate limit on performance.

\section{Information Theory for Very Large MIMO Arrays}
\label{sec:it}
Shannon's information theory provides, under very precisely specified
conditions, bounds on attainable performance of communications
systems. According to the noisy-channel coding theorem, for any
communication link there is a \emph{capacity} or \emph{achievable
  rate}, such that for any transmission
rate less than the capacity, there exists a coding scheme that makes
the error-rate arbitrarily small. 

The classical point-to-point MIMO link begins our discussion
and it serves to highlight the limitations of systems in which the
working antennas are compactly clustered at both ends of the
link. This leads naturally into the topic of multi-user MIMO which is
where we envision very large MIMO will show its greatest utility. The
Shannon theory simplifies greatly for
large numbers of antennas  and it
suggests capacity-approaching strategies.

\subsection{Point-to-point MIMO}

\subsubsection{Channel model}

A point-to-point MIMO link consists of a transmitter having an array
of $n_\mathrm{t}$ antennas, a receiver having an array of
$n_\mathrm{r}$ antennas, with both arrays  connected by a channel such
that every receive antenna is subject to the combined action of all
transmit antennas. The simplest narrowband memoryless channel has the
following mathematical description; for each use of the channel we have
\begin{equation}
\mathbf{x} = \sqrt{\rho}\vec{G} \mathbf{s} + \mathbf{w} \ ,
\label{eq:sig_p2p}
\end{equation}
where $\mathbf{s}$ is the $n_\mathrm{t}$-component vector of
transmitted signals, $\mathbf{x}$ is the $n_\mathrm{r}$-component vector of
received signals, $\vec{G}$ is the $n_\mathrm{r}
\times n_\mathrm{t}$ propagation matrix of complex-valued channel
coefficients, and 
$\mathbf{w}$ is the $n_\mathrm{r}$-component vector of receiver noise. The
scalar $\rho$ is a measure of the Signal-to-Noise Ratio (SNR) of the
link: it is proportional to the transmitted power divided by the
noise-variance, and it also absorbs various normalizing constants. In
what follows we assume a normalization such that the expected total
transmit power is unity,
\begin{equation}
\mathrm{E} \left\{ \|\vec{s}\|^2 \right\} =1 \ ,
\label{eq:power_constraint}
\end{equation}
where the components of the additive noise vector are Independent and
Identically Distributed (IID)
zero-mean and unit-variance circulary-symmetric complex-Gaussian random variables ($\mathcal{CN}(0,1)$). Hence if
there were only one antenna at each end of the link, then within
(\ref{eq:sig_p2p}) the quantities $\mathbf{s}$, $\vec{G}$, $\mathbf{x}$ and
$\mathbf{w}$ would be scalars, and the SNR would be equal to $\rho |\vec{G}|^2$.

In the case of a wide-band, frequency-dependent (``delay-spread'')
channel, the channel is described by a matrix-valued impulse response
or by the equivalent matrix-valued frequency response. One may
conceptually decompose the channel into parallel 
independent narrow-band channels, each of which is described in the
manner of (\ref{eq:sig_p2p}). Indeed, Orthogonal Frequency-Division
Multiplexing (OFDM) rigorously performs this decomposition.

\subsubsection{Achievable rate}

With IID complex-Gaussian inputs, the (instantaneous) mutual information between the
input and the output of the point-to-point MIMO channel
(\ref{eq:sig_p2p}), \emph{under the assumption that the receiver has
  perfect knowledge of the channel matrix, $\vec{G}$}, measured in
bits-per-symbol (or equivalently bits-per-channel-use) is 
\begin{equation}
C = I(\vec{x};\vec{s})=\log_2 \det \left(\vec{I}_{n_\mathrm{r}} + \frac{\rho}{n_\mathrm{t}} \vec{G}
  \vec{G}\HT \right) \ ,
\label{eq:C_p2p}
\end{equation} 
where $I(\vec{x};\vec{s})$ denotes the mutual information operator, $\vec{I}_{n_\mathrm{r}}$ denotes the $n_\mathrm{r} \times
n_\mathrm{r}$ identity matrix 
and 
 the superscript ``$\mathrm{H}$'' denotes the Hermitian transpose
 \cite{Foschini1999}. The actual \emph{capacity} of the channel
 results if the inputs are optimized according to the water-filling principle. In
 the case that $\vec{G}\vec{G}\HT$ equals a scaled identity matrix,
 $C$ is in fact the capacity.

To approach the achievable rate $C$, the transmitter does not have to know the channel, however it must be informed of the numerical value of the achievable rate. Alternatively, if the channel is governed by known statistics,
then the transmitter can set a rate which is consistent with an
acceptable \emph{outage probability}.  For the special case of one
antenna at each end of the link, the achievable rate (\ref{eq:C_p2p}) becomes
that of the scalar additive complex Gaussian noise channel,
\begin{equation}
C = \log_2 \left( 1 + \rho |\vec{G}|^2 \right) \ .
\label{eq:C_scalar}
\end{equation}

The implications of (\ref{eq:C_p2p}) are most easily seen by
expressing the achievable rate in terms of the singular values of the propagation
matrix,
\begin{equation}
\vec{G} = \vec{\Phi}\vec{\mathbf{D}}_\mathbf{\nu} \vec{\Psi}\HT \ ,
\label{eq:svd}
\end{equation}
where $\vec{\Phi}$ and $\vec{\Psi}$ are unitary matrices of dimension
$n_\mathrm{r} \times n_\mathrm{r}$ and $n_\mathrm{t} \times
n_\mathrm{t}$ respectively, and $\vec{D}_\mathbf{\nu}$ is a $n_\mathrm{r}
\times n_\mathrm{t}$ diagonal matrix whose diagonal elements are the
singular values, $\{ \nu_1, \ \nu_2, \ \cdots \ \nu_{\min
  (n_\mathrm{t},n_\mathrm{r})} \} $. The achievable rate (\ref{eq:C_p2p}),
expressed in terms of the singular values, 
\begin{equation}
C = \sum_{\ell = 1}^{\min (n_\mathrm{t},n_\mathrm{r})} \log_2 \left( 1 +
\frac{\rho \nu_\ell^2}{n_\mathrm{t}} \right) \ ,
\label{eq:C_p2p_sv}
\end{equation}
is equivalent to the combined achievable rate of parallel links for which the
$\ell$-th link has an SNR of $\rho \nu_\ell^2 / n_\mathrm{t}$. With
respect to the achievable rate, it is
interesting to consider the best and the worst possible distribution of
singular values. Subject to the constraint (obtained directly from
(\ref{eq:svd})) that
\begin{equation}
\sum_{\ell = 1}^{\min (n_\mathrm{t},n_\mathrm{r})} \nu_\ell^2 = \rm{Tr} \left(
\vec{G} \vec{G}\HT \right) \ ,
\label{eq:sv_constraint}
\end{equation}
where ``$ \mathrm{Tr}$'' denotes ``trace'', the worst case is when all
but one of the singular values are equal to 
zero, and the best case is when all of the $\min
(n_\mathrm{t},n_\mathrm{r})$ singular values are equal (this is a
simple consequence of the concavity of the logarithm). The two cases
bound the achievable rate (\ref{eq:C_p2p_sv}) as follows,
\begin{equation}
\log_2 \left(1 + \frac{\rho \cdot \mathrm{Tr} \left(\vec{G}\vec{G}\HT
  \right)}{n_\mathrm{t}} \right) 
\leq C \leq \min (n_\mathrm{t},n_\mathrm{r}) \cdot \log_2 \left( 1 +
\frac{\rho \cdot \mathrm{Tr} \left( \vec{G}\vec{G}\HT \right)}{n_\mathrm{t} \min
(n_\mathrm{t},n_\mathrm{r})} \right) \ .
\label{eq:C_p2p_bound1}
\end{equation}
If we assume that a normalization has been performed such that the
magnitude of a propagation coefficient is 
typically equal to one, then $\mathrm{Tr} \left( \vec{G}\vec{G}\HT \right)
\approx n_\mathrm{t} 
n_\mathrm{r}$, and the above bounds simplify as follows,
\begin{equation}
\log_2 \left(1 + \rho n_\mathrm{r} \right)
\leq C \leq \min (n_\mathrm{t},n_\mathrm{r})\cdot \log_2 \left( 1 +
\frac{\rho \max (n_\mathrm{t},n_\mathrm{r})}{n_\mathrm{t} } \right) \ .
\label{eq:C_p2p_bound2}
\end{equation}
The rank-1 (worst) case occurs either for compact arrays under Line-of-Sight (LOS) propagation conditions such that the transmit array cannot resolve individual elements of the receive array and vice-versa, or under extreme keyhole propagation conditions. The equal singular value (best) case is
approached when the entries of the propagation matrix are IID random variables. Under favorable propagation
conditions and a high SNR, the achievable rate is proportional to the smaller
of the number of transmit and receive antennas.

\subsubsection{Limiting cases}
Low SNRs can be
experienced by terminals at the edge of a cell.
For low SNRs only beamforming gains are important and the achievable rate (\ref{eq:C_p2p}) becomes
\begin{eqnarray}
C_{\rho \rightarrow 0} & \approx & \frac{\rho \cdot
  \mathrm{Tr} \left(\vec{G}\vec{G}\HT\right)}{n_\mathrm{t} \ln 2} \nonumber \\
& \approx & \frac{\rho n_\mathrm{r}}{\ln 2} \ .
\label{eq:C_p2p_rho0}
\end{eqnarray}
This expression is independent of $n_t$, and thus, even under the most favorable propagation conditions the
multiplexing gains are lost, and from the perspective of achievable rate,
multiple transmit antennas are of no value. 

Next let the number of transmit antennas grow large while keeping the
number of receive antennas constant. We furthermore assume that the
row-vectors of the propagation matrix are asymptotically
orthogonal. As a consequence \cite{Mattaiou10}
\begin{equation}
\left( \frac{\vec{G} \vec{G}\HT}{n_\mathrm{t}} \right)_{n_\mathrm{t} \gg
  n_\mathrm{r}} \approx 
\vec{I}_{n_\mathrm{r}} \ ,
\label{eq:G_r_orthog}
\end{equation}
and the achievable rate (\ref{eq:C_p2p}) becomes
\begin{eqnarray}
C_{n_\mathrm{t} \gg n_\mathrm{r}} & \approx & \log_2 \det \left(
\vec{I}_{n_\mathrm{r}} + \rho \cdot \vec{I}_{n_\mathrm{r}}  \right) \nonumber \\
& = & n_\mathrm{r} \cdot \log_2 (1 + \rho) \ ,
\label{eq:C_p2p_large_n_t}
\end{eqnarray}
which matches the upper bound (\ref{eq:C_p2p_bound2}). 

Then, let the number of receive antennas grow large while keeping the
number of transmit antennas constant. We also assume that the
column-vectors of the propagation matrix are asymptotically
orthogonal, so
\begin{equation}
\left( \frac{\vec{G}\HT \vec{G}}{n_\mathrm{r}} \right) _{n_\mathrm{r} \gg
  n_\mathrm{t}} \approx \vec{I}_{n_\mathrm{t}} \ .
\label{eq:G_t_orthog}
\end{equation}
The identity $\det (I + \vec{A}\vec{A}\HT) = \det(I + \vec{A}\HT \vec{A})$, combined with
(\ref{eq:C_p2p}) and (\ref{eq:G_t_orthog}), yields
\begin{eqnarray}
C_{n_\mathrm{r} \gg n_\mathrm{t}} & = & \log_2 \det \left(
\vec{I}_{n_\mathrm{t}} + \frac{\rho}{n_\mathrm{t}} \vec{G}\HT \vec{G} \right) \nonumber \\
& \approx & n_\mathrm{t} \cdot \log_2 \left( 1 + \frac{\rho
  n_\mathrm{r}}{n_\mathrm{t}} \right) \ ,
\label{eq:C_p2p_large_n_r}
\end{eqnarray}
which again matches  the upper bound (\ref{eq:C_p2p_bound2}). So an
excess number of transmit or receive antennas, combined with
asymptotic orthogonality of the propagation vectors, constitutes a
highly desirable scenario. Extra
receive antennas continue to boost the effective SNR, and could in
theory compensate for a low SNR and restore multiplexing gains which
would otherwise be lost as in (\ref{eq:C_p2p_rho0}). Furthermore,
orthogonality of the propagation vectors implies that IID
complex-Gaussian inputs are optimal so that the achievable rates
(\ref{eq:G_t_orthog}) and (\ref{eq:C_p2p_large_n_r}) are in fact
the true channel capacities.

\subsection{Multi-user MIMO}

The attractive multiplexing gains promised by point-to-point MIMO
require a favorable propagation environment and a good
SNR. Disappointing performance can occur in LOS propagation
or when the terminal is at the edge of the cell. Extra
receive antennas can compensate for a low SNR, but for the forward
link this adds to the complication and expense of the
terminal. Very large MIMO can fully address the shortcomings of
point-to-point MIMO.

If we split up the antenna array at one end of a point-to-point MIMO
link into autonomous antennas we obtain the qualitatively different
Multi-User MIMO (MU-MIMO). Our context for discussing this is an array of $M$
antennas - for example a base station - which simultaneously serves
$K$ autonomous terminals. (Since we want to study both forward- and
reverse link transmission, we now abandon the notation $n_\mathrm{t}$
and $n_\mathrm{r}$.) In what follows we assume that each terminal
has only one antenna. Multi-user MIMO differs from point-to-point MIMO
in two respects: first, the terminals are typically separated by many
wavelengths, and second, the terminals cannot collaborate among
themselves, either to transmit or to receive data.

\subsubsection{Propagation} \label{InfThPropagation}
We will assume TDD operation, so the reverse link
propagation matrix is merely the transpose of the forward link 
propagation matrix. Our emphasis on TDD rather than FDD is driven by the need to acquire channel state-information between extreme numbers of service antennas and much smaller numbers of terminals. The time required to transmit reverse-link pilots is independent of the number of  antennas, while the time required to transmit forward-link pilots is proportional to the number of antennas. 
The propagation matrix in the reverse link, $\vec{G}$, dimensioned $M \times K$, is the
product of a $M \times K$ matrix, $\vec{H}$, which accounts for small scale fading
(i.e., which changes over intervals of a wavelength or less), and a $K
\times K$ diagonal matrix, $\vec{D}_\mathbf{\beta}^{1/2}$, whose diagonal
elements constitute a $K \times 1$ vector, $\vec{\beta}$, of large scale fading
coefficients,
\begin{equation}
\vec{G} = \vec{H} \vec{D}_\mathbf{\beta}^{1/2} .
\label{eq:G_factor}
\end{equation}
 The large scale fading accounts for path loss and
shadow fading. Thus the $k$-th column-vector of $\vec{H}$ describes the
small scale
fading between the $k$-th terminal and the $M$ antennas, while the
$k$-th diagonal element of $\vec{D}_\mathbf{\beta}^{1/2}$ is the large scale fading
coefficient. By assumption, the antenna array is sufficiently compact
that all of the propagation paths for a particular terminal are
subject to the same large scale fading. We normalize the large scale fading
coefficients such that the small scale fading coefficients typically have
magnitudes of one.

For multi-user MIMO with large arrays, the number of antennas greatly exceeds
the number of terminals. Under the most favorable propagation
conditions the column-vectors of the propagation matrix are
asymptotically orthogonal,
\begin{eqnarray}
\left( \frac{\vec{G}\HT \vec{G}}{M} \right)_{M \gg K} & = & \vec{D}_{\mathbf{\beta}}^{1/2}
\left( \frac{ \vec{H}\HT \vec{H}}{M} \right)_{M \gg K} \vec{D}_{\mathbf{\beta}}^{1/2}
\nonumber \\ 
& \approx & \vec{D}_{\mathbf{\beta}} .
\label{eq:G_mu_orthog}
\end{eqnarray}

\subsubsection{Reverse link}
\label{sec:reverse_capacity}
On the reverse link, for each channel use, the $K$ terminals
collectively transmit a $K \times 1$ vector of QAM symbols,
$\mathbf{q}_\mathrm{r}$, and the antenna array receives a $M \times 1$
vector, $\mathbf{x}_\mathrm{r}$,
\begin{equation}
\mathbf{x}_\mathrm{r} = \sqrt{\rho_\mathrm{r}} \vec{G} \mathbf{q}_\mathrm{r} +
\mathbf{w}_\mathrm{r} \ ,
\label{eq:sig_rev}
\end{equation}
where $\mathbf{w}_\mathrm{r}$ is the $M \times 1$ vector of receiver
noise whose components are independent and distributed as $\mathcal{CN}(0,1)$. The quantity
$\rho_\mathrm{r}$ is proportional to the ratio of power divided by
noise-variance. Each
terminal is constrained to have an expected power of one,
\begin{equation}
\mathrm{E} \left\{ |q_{\mathrm{r} k}|^2  \right\} = 1, \ k=1, \cdots , K \ .
\label{eq:power_rev}
\end{equation}
We assume that the base station knows the channel.

Remarkably, the total throughput (e.g., the achievable sum-rate) of
reverse link multi-user MIMO is no less than if the terminals could
collaborate among themselves \cite{tv:05},
\begin{equation}
C_\mathrm{sum \_ r} = \log_2 \det \left( \vec{I}_K + \rho_\mathrm{r} \vec{G}\HT
\vec{G}  \right) \ . 
\label{eq:C_rev}
\end{equation}
If collaboration were possible it could definitely make channel coding
and decoding easier, but it would not alter the ultimate
sum-rate. The sum-rate is not generally shared equally by the
terminals; consider for example the case where the slow fading
coefficient is near-zero for some terminal.

Under favorable propagation conditions (\ref{eq:G_mu_orthog}), if
there is a large number of antennas compared with terminals, then the
asymptotic sum-rate is
\begin{eqnarray}
{C_\mathrm{sum \_ r}}_{M \gg K} & \approx & \log_2 \det \left( \vec{I}_K + M
\rho_\mathrm{r} \vec{D}_\beta \right) \nonumber \\
& = & \sum_{k=1}^K \log_2 \left( 1 + M
\rho_\mathrm{r} \beta_k \right) \ .
\label{eq:C_rev_large}
\end{eqnarray}
This has a nice intuitive interpretation if we
assume that the columns of the propagation matrix are nearly
orthogonal, i.e., 
$\vec{G}\HT \vec{G} \approx M \cdot \vec{D}_\beta$. Under this assumption, the base station could
process its received signal by a Matched-Filter (MF),
\begin{eqnarray}
\vec{G}\HT \mathbf{x}_\mathrm{r} & = & \sqrt{\rho_\mathrm{r}} \vec{G}\HT \vec{G}
\mathbf{q}_\mathrm{r} + \vec{G}\HT \mathbf{w}_\mathrm{r} \nonumber \\
& \approx & M \sqrt{ \rho_\mathrm{r}} \vec{D}_{\mathbf{\beta}}
\mathbf{q}_\mathrm{r} + \vec{G}\HT \mathbf{w}_\mathrm{r} \ .
\label{eq:rev_proc}
\end{eqnarray}
This processing separates the signals transmitted by the
different terminals. The decoding of the transmission from the $k$-th terminal
requires only the $k$-th component of (\ref{eq:rev_proc}); this has an
SNR of $M \rho_\mathrm{r} \beta_k$, which in turn yields an individual
rate for that terminal,  corresponding to the $k$-th term in the
sum-rate (\ref{eq:C_rev_large}). 
 
\subsubsection{Forward link}
For each use of the channel the base station transmits a $M \times 1$
vector, $\mathbf{s}_\mathrm{f}$, through its $M$ antennas, and the $K$
terminals collectively receive a $K \times 1$ vector,
$\mathbf{x}_\mathrm{f}$,
\begin{equation}
\mathbf{x}_\mathrm{f} = \sqrt{\rho_\mathrm{f}} \vec{G}^\mathrm{T}
\mathbf{s}_\mathrm{f} + \mathbf{w}_\mathrm{f} \ ,
\label{eq:sig_for}
\end{equation}
where the superscript ``T'' denotes ``transpose'', and
$\mathbf{w}_\mathrm{f}$ is the $K \times 1$ vector of receiver 
noise whose components are independent and distributed as $\mathcal{CN}(0,1)$. The
quantity $\rho_\mathrm{f}$ is proportional to the ratio of power to
noise-variance. The
total transmit power is independent of the number of antennas,
\begin{equation}
\mathrm{E} \left\{ \| \mathbf{s}_\mathrm{f}\|^2 \right\} = 1 \ .
\label{eq:power_for}
\end{equation}

The known capacity result for this channel, see e.g. \cite{Jindal03,WSS06},  assumes that the
terminals as well as the base station know the channel. Let
$\vec{D}_{\gamma}$ be a diagonal matrix
 whose diagonal elements constitute a $K\times 1$ vector $\vec{\gamma}$.
To obtain the sum-capacity requires performing a constrained optimization,
\begin{eqnarray}
C_\mathrm{sum \_ f} & = & \max_{\{\gamma_k\}} \log_2 \det
\left( \vec{I}_M + \rho_\mathrm{f} \vec{G} \vec{D}_\mathbf{\gamma} \vec{G}\HT
\right), \nonumber \\
& \mathrm{subject \ to} &
 \sum_{k=1}^K \gamma_k = 1, \ \gamma_k \geq 0,  \, \forall \ k \ .
\label{eq:C_sum_for}
\end{eqnarray}

Under favorable propagation conditions (\ref{eq:G_mu_orthog}) and a
large excess of antennas, the sum-capacity has a simple asymptotic
form,
\begin{eqnarray}
{C_\mathrm{sum \_ f}}_{M \gg K} & = & \max_{\{\gamma_k\}}
\log_2 \det \left( \vec{I}_K + \rho_\mathrm{f} \vec{D}_{\mathbf{\gamma}}^{1/2}
\vec{G}\HT \vec{G} \vec{D}_{\mathbf{\gamma}}^{1/2}  \right) \nonumber \\
& \approx & \max_{\{\gamma_k\}} \log_2 \det \left( \vec{I}_K +
M \rho_\mathrm{f} \vec{D}_\mathbf{\gamma} \vec{D}_\mathbf{\beta}
\right) \nonumber \\
& = & \max_{\{\gamma_k\}} \sum_{k=1}^K \log_2
\left( 1 + M \rho_\mathrm{f} \gamma_k \beta_k  \right)  \ ,
\label{eq:C_sum_for_big}
\end{eqnarray}
where $\mathbf{\gamma}$ is constrained as in
(\ref{eq:C_sum_for}). This result makes intuitive sense if the columns
of the propagation matrix are nearly 
orthogonal which occurs asymptotically as the number of antennas grows. Then the transmitter could use a simple 
MF linear precoder,
\begin{equation}
\mathbf{s}_\mathrm{f} = \frac{1}{\sqrt{M}} \vec{G}^*
\vec{D}_\mathbf{\beta}^{-1/2} \vec{D}_\mathrm{p}^{1/2}
\mathbf{q}_\mathrm{f} ,
\label{eq:mf_precoder}
\end{equation}
where $\mathbf{q}_\mathrm{f}$ is the vector of QAM symbols intended
for the terminals such that $\mathrm{E} \left\{| q_{\mathrm{f}k}|^2 =
1 \right\}$, and $\mathbf{p}$ is a vector of powers such that
$\sum_{k=1}^K p_k = 1$. The substitution of (\ref{eq:mf_precoder})
into (\ref{eq:sig_for}) yields the following,
\begin{equation}
\mathbf{x}_\mathrm{f} \approx \sqrt{\rho_\mathrm{f} M}
\vec{D}_\mathbf{\beta}^{1/2} \vec{D}_\mathbf{p}^{1/2}
\mathbf{q}_\mathrm{f} + \mathbf{w}_\mathrm{f} ,
\label{eq:x_mf_precode}
\end{equation}
which yields an achievable sum-rate of
$\sum_{k=1}^K \log_2
\left( 1 + M \rho_\mathrm{f} p_k \beta_k  \right)$
- identical to the sum-capacity (\ref{eq:C_sum_for_big}) if we identify
  $\mathbf{p} = \vec{\gamma}$.

\section{Antenna and propagation aspects of Very Large MIMO}

The performance of all types of MIMO systems strongly depends on
properties of the antenna arrays and the propagation environment in
which the system is operating. The complexity of the propagation
environment, in combination with the capability of the antenna arrays
to exploit this complexity, limits the achievable system
performance. When the number of antenna elements in the arrays
increases, we meet both opportunities and challenges. The
opportunities include increased capabilities of exploiting the
propagation channel, with better spatial resolution. With well
separated ideal antenna elements, in a sufficiently complex
propagation environment and without directivity and mutual coupling,
each additional antenna element in the array adds another degree of
freedom that can be used by the system. In reality, though, the
antenna elements are never ideal, they are not always well separated,
and the propagation environment may not be complex enough to offer the
large number of degrees of freedom that a large antenna array could
exploit. In this section we illustrate and discuss some of these
opportunities and challenges, starting with an example of how more
antennas in an ideal situation improves our capability to focus the
field strength to a specific geographical point (a certain user). This
is followed by an analysis of how realistic (non-ideal) antenna arrays
influence the system performance in an ideal propagation
environment. Finally, we use channel measurements to address
properties of a real case with a 128-element base station array
serving 6 single-antenna users.

\subsection{Spatial focus with more antennas}

Precoding of an antenna array is often said to \emph{direct} the
signal from the antenna array towards one or more receivers. In a pure
LOS environment, directing means that the antenna array forms a beam
towards the intended receiver with an increased field strength in a
certain direction from the transmitting array. In propagation
environments where non-LOS components dominate, the concept of
directing the antenna array towards a certain receiver becomes more
complicated. In fact, the field strength is not necessarily focused in
the \emph{direction} of the intended receiver, but rather to a
geographical \emph{point} where the incoming multipath components add
up constructively. Different techniques for focusing transmitted energy
to a specific location have been addressed in several contexts. In
particular, it has drawn attention in the form of Time Reversal (TR)
where the transmitted signal is a time-reversed replica of the channel
impulse response. TR with single
as well as multiple antennas has been demonstrated lately in, e.g.,
\cite{LRT2004,HeB2004}. In the context of this paper the most
interesting case is MISO, and here we speak of Time-Reversal
Beam Forming (TRBF). While most communications applications of TRBF
address a relatively small number of antennas, the same basic
techniques have been studied for almost two decades in medical
extracorporeal lithotripsy applications \cite{TWF1996} with a large
number of ``antennas'' (transducers).

\begin{figure}
        \centering
                \includegraphics[width=0.7\columnwidth]{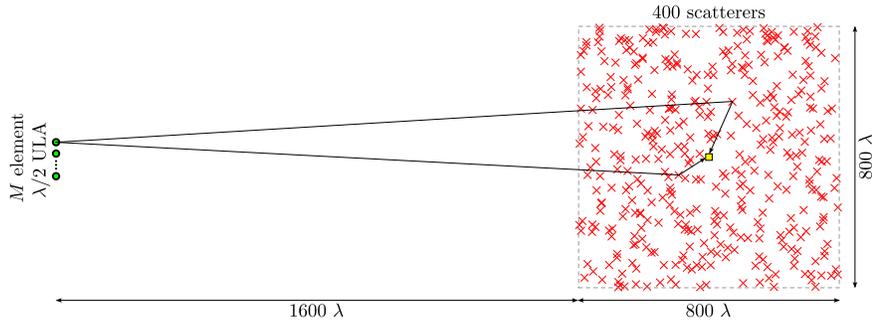}
        \caption{Geometry of the simulated dense scattering
          environment, with 400 uniformly distributed scatterers in a
          $800 \times 800$ $\lambda$ area. The transmit $M$-element
           ULA is placed at a distance of 1600 $\lambda$ from
          the edge of the scatterer area with its broadside pointing
          towards the center. Two single scattering paths from the
          first ULA element to an intended receiver in the center of
          the scatterer area are shown.}
        \label{fig:spatialfocus_geometry}
\end{figure}

To illustrate how large antenna arrays can focus the electromagnetic
field to a certain geographic point, even in a narrowband channel, we
use the simple geometrical channel model shown in
Figure~\ref{fig:spatialfocus_geometry}. The channel is composed of 400
uniformly distributed scatterers in a square of dimension $800 \lambda
\times 800 \lambda$, where $\lambda$ is the signal wavelength. The scattering points ($\times$) shown in the
figure are the actual ones used in the example below. The broadside
direction of the $M$-element Uniform Linear Array (ULA) with adjacent element spacing of $d=\lambda/2$ is
pointing towards the center of the scatterer area. Each
single-scattering multipath component is subject to an inverse
power-law attenuation, proportional to distance squared (propagation
exponent 2), and a random reflection coefficient with IID complex Gaussian distribution (giving a Rayleigh distributed amplitude and a uniformly distributed phase). This model creates a 
field strength that varies rapidly over the geographical area,
typical of small-scale fading. With a complex enough scattering
environment and a sufficiently large element spacing in the transmit
array, the field strength resulting from different elements in the
transmit array can be seen as independent.

\begin{figure}
        \centering
                \includegraphics[width=0.6\columnwidth]{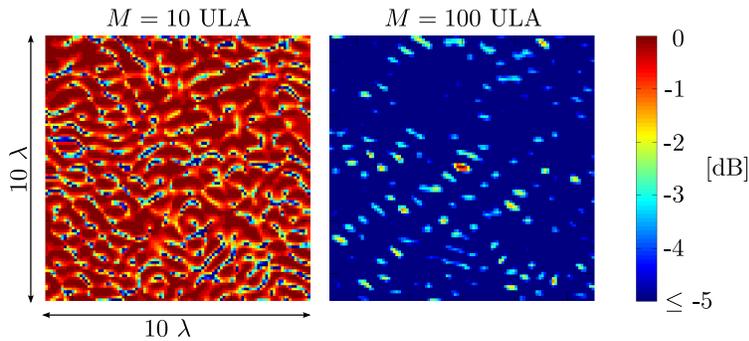}
        \caption{Normalized fieldstrength in a $10 \times 10$
                $\lambda$ area centered around the receiver to which
                the beamforming is done. The left and right pseudo
                color plots show the field strength when an $M=10$ and an $M=100$ ULA are used together with MF precoding to focus the signal to a receiver in the center of the area.}
        \label{fig:spatialfocus_field}
\end{figure}

In Figure~\ref{fig:spatialfocus_field} we show the resulting
normalized field strength in a small $10 \lambda \times 10 \lambda$
environment around the receiver to which we focus the transmitted
signal (using MF precoding), for ULAs with $d=\lambda/2$ of size
$M=10$ and $M=100$ elements. The normalized field strength shows how
much weaker the field strength is in a certain position when the
spatial signature to the center point is used rather than the correct
spatial signature for that point. Hence, the normalized field strength
is 0\ dB at the center of both figures, and negative at all other
points. Figure~\ref{fig:spatialfocus_field} illustrates two important
properties of the spatial MF precoding: (i) that the field strength
can be focused to a point rather than in a certain direction and (ii)
that more antennas improve the ability to focus energy to a certain
point, which leads to less interference between spatially separated
users. With $M=10$ antenna elements, the focusing of the field
strength is quite poor with many peaks inside the studied
area. Increasing $M$ to 100 antenna elements, for the same propagation
environment, considerably improves the field strength focusing and it
is more than 5 dB down in most of the studied area.

While the example above only illustrates spatial MF precoding in the
narrowband case, the TRBF techniques exploit both the spatial and
temporal domains to achieve an even stronger spatial focusing of the
field strength. With enough many antennas and favorable propagation
conditions, TRBF will not only focus power and yield a high spectral
efficiency through spatial multiplexing to many terminals. It will
also reduce, or in the ideal case completely eliminate, inter-symbol interference. In other words, one could
dispense with OFDM and its redundant cyclic prefix. Each base station
antenna would 1) merely convolve the data sequence intended for the $k$-th
terminal with the conjugated, time-reversed version of his estimate for
the channel impulse response to the $k$-th terminal, 2) sum the $K$
convolutions, and 3) feed that sum into his antenna. Again, under
favorable propagation conditions, and a large number of antennas,
inter-symbol interference will decrease significantly.

\subsection{Antenna aspects}

It is common within the signal processing, communications and
information theory communities to assume that the transmit and receive
antennas are isotropic and uni-polarized electromagnetic wave
radiators and sensors, respectively. In reality, such isotropic
unipolar antennas do not exist, according to fundamental laws of
electromagnetics. Non-isotropic antenna patterns will influence the
MIMO performance by changing the spatial correlation. For example, directive 
antennas pointing in distinct directions tend to experience a lower correlation 
than non-directive antennas, since each of these directive antennas ``see'' signals 
arriving from a distinct angular sector.   

In the context of an array of antennas, it is also common in these communities to assume
that there is no electromagnetic interaction (or mutual coupling)
among the antenna elements neither in the transmit nor in the receive
mode. This assumption is only valid when the antennas are well
separated from one another.


In the rest of this section we consider very large MIMO arrays where
the overall aperture of the array is constrained, for example, by the
size of the supporting structure or by aesthetic
considerations. Increasing the number of antenna elements implies that
the antenna separation decreases. This problem has been examined in
recent papers, although the focus is often on spatial correlation and
the effect of coupling is often neglected, as
in~\cite{PoAK03,WeGJ05,HaFu06}. In \cite{Jana02}, the effect of
coupling on the capacity of fixed length ULAs is studied. In general,
it is found that mutual coupling has a substantial impact on capacity
as the number of antennas is increased for a fixed array aperture.

It is conceivable that the capacity performance in
\cite{Jana02} can be improved by compensating for the effect of mutual
coupling. Indeed, coupling compensation is a topic of current
interest, much driven by the desire of implementing MIMO arrays in a
compact volume, such as mobile terminals (see \cite{Lau10} and
references therein). One interesting result is that coupling among
co-polarized antennas can be perfectly mitigated by the use of optimal multiport
impedance matching radio frequency circuits \cite{WaJe04a}. This
technique has been experimentally demonstrated only for up to four
antennas, though in principle it can be applied to very large MIMO
arrays \cite{VoWS08}. Nevertheless, the effective cancellation of
coupling also brings about diminishing bandwidth in one or more output ports 
as the antenna spacing decreases \cite{LaAK06b}. This can be understood 
intuitively in that, in the limit of small antenna spacing, the array effectively 
reduces to only one antenna. Thus, one can only expect the array to offer the 
same characteristics as a single antenna. Furthermore, implementing 
practical matching circuits will introduce ohmic losses, which reduces the gain
that is achievable from coupling cancellation~\cite{Lau10}.


Another issue to consider is that due to the constraint in array
aperture, very large MIMO arrays are expected to be implemented in a
2D or 3D array structure, instead of as a linear array as in \cite{Jana02}.
A linear array with antenna elements of identical gain
patterns (e.g., isotropic elements) suffers from the problem of
front-back ambiguity, and is also unable to resolve signal paths in
\textit{both} azimuth and elevation. However, one drawback of having a
dense array implementation in 2D or 3D is the increase of coupling
effects due to the increase in the number of adjacent antennas. For
the square array (2D) case, there are up to four adjacent antennas (located
at the same distance) for each antenna element, and in 3D there are up
to 6. A further problem that is specific to 3D arrays is that only the
antennas located on the surface of the 3D array contribute to
the information capacity~\cite{MoBB00}, which in effect restricts the
usefulness of dense 3D array implementations. This 
is a consequence of the integral representation of Maxwell's
equations, by which the electromagnetic field inside the volume of the
3D array is fully described by the field on its surface (assuming
sufficiently dense sampling), and therefore no additional information
can be extracted from elements inside the 3D array.

Moreover, in outdoor cellular environments, signals tend to arrive 
within a narrow range of elevation angles. Therefore, it may not be 
feasible for the antenna system to take advantage of the resolution 
in elevation offered by dense 2D or 3D arrays to perform signaling 
in the vertical dimension.

\begin{figure}
         \centering
                 \includegraphics[width=0.7\columnwidth]{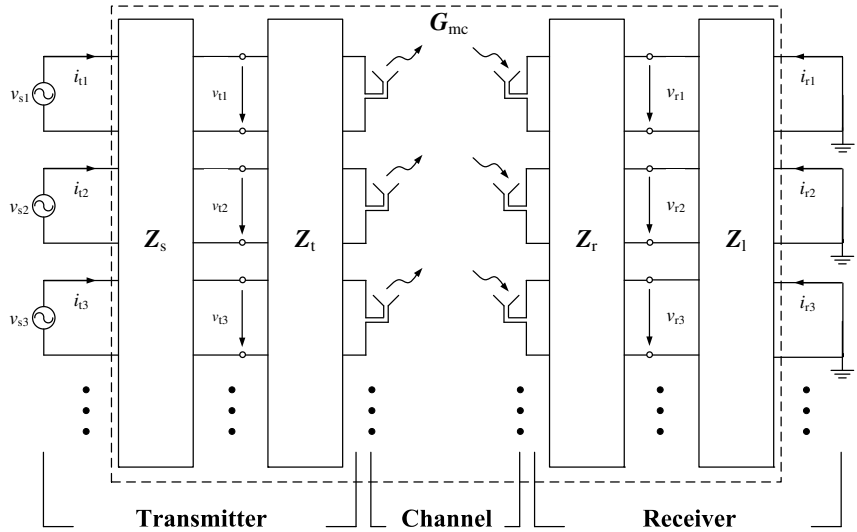}
         \caption{Diagram of a MIMO system with antenna impedance matrices and matching networks at both link ends (freely reproduced from \cite{FeFL08}).}
         \label{fig:coupling_model}
 \end{figure}

The complete Single-User MIMO (SU-MIMO) signal model with antennas and
matching circuit in Figure \ref{fig:coupling_model} (reproduced from
\cite{FeFL08}) is used to demonstrate the performance degradation
resulting from correlation and mutual coupling in very large arrays
with fixed apertures. In the figure, ${\vec{Z}}_\mathrm{t}$ and
${\vec{Z}}_\mathrm{r}$ are the impedance matrices of the transmit and
receive arrays, respectively, $i_{\mathrm{t}i}$ and $i_{\mathrm{r}i}$
are the excitation and received currents (at the $i$-th port) of the
transmit and receive systems, respectively, and $v_{\mathrm{s}i}$ and
$v_{\mathrm{r}i}$ (${\vec{Z}}_\mathrm{s}$ and ${\vec{Z}}_\mathrm{l}$)
are the source and load voltages (impedances), respectively, and
$v_{\mathrm{t}i}$ is the terminal voltage across the $i$-th transmit
antenna port. $\vec{G}_\mathrm{mc}$ is the \textit{overall} channel of
the system, including the effects of antenna coupling and matching
circuits.

Recall that the instantaneous capacity\footnote{From this
  point and onwards, we shall for simplicity refer to the $\log-\det$
  formula with IID complex-Gaussian inputs as ``the capacity'' to
  avoid the more clumsy notation of ``achievable rate''.} is given by (\ref{eq:C_p2p}) and equals
\begin{equation}
C_\mathrm{mc}=\log_2\det\left(\vec{I}_{n}+\frac{\rho}{n_\mathrm{t}}{\hat{\vec{G}}}_\mathrm{mc}{\hat{\vec{G}}}_\mathrm{mc}\HT \right) ,
\label{eq:capacity_coupled}
\end{equation}
where
\begin{equation}
{\hat{\vec{G}}}_\mathrm{mc}=2r_{11} \vec{R}_\mathrm{l}^{1/2}({\vec{Z}}_\mathrm{l}+{\vec{Z}}_\mathrm{r})^{-1}{\vec{G}}{\vec{R}}_\mathrm{t}^{-1/2} ,
\label{eq:G_coupled}
\end{equation}
is the overall MIMO channel based on the complete SU-MIMO signal model, 
$\vec{G}$ represents the propagation channel as seen by the transmit and receive antennas, and
${\vec{R}}_\mathrm{l}=\Re\left\{{\vec{Z}}_\mathrm{l}\right\}$,
${\vec{R}}_\mathrm{t}=\Re\left\{{\vec{Z}}_\mathrm{t}\right\}$. Note
that ${\hat{\vec{G}}}_\mathrm{mc}$ is the \textit{normalized} version
of $\vec{G}_\mathrm{mc}$ shown in Figure~\ref{fig:coupling_model},
where the normalization is performed with respect to the average
channel gain of a SISO system~\cite{FeFL08}. The source impedance matrix $\vec{Z}_\mathrm{s}$ does
not appear in the expression, since ${\hat{\vec{G}}}_\mathrm{mc}$
represents the transfer function between the transmit and receive
power waves, and $\vec{Z}_\mathrm{s}$ is implicit in $\rho$ \cite{FeFL08}. 


To give an intuitive feel for the effects of mutual coupling, we next provide two examples
of the impedance matrix ${\vec{Z}}_\mathrm{r}$\footnote{For a given antenna array, ${\vec{Z}}_\mathrm{t}={\vec{Z}}_\mathrm{r}$ by the
principle of reciprocity.}, one for small adjacent antenna spacing (0.05$\lambda$)
and one for moderate spacing (0.5$\lambda$). 
The following numerical values  are obtained from the induced electromotive force  method~\cite{Bal05} for a ULA consisting of three parallel dipole antennas:
\begin{equation}\label{eq:imp_matrix1}
\vec{Z}_\mathrm{r}(0.05\lambda)=\left[ \begin{array}{ccc} 72.9 +j42.4 & 71.4+j24.3 & 67.1+j7.6 \\ 71.4 +j24.3 & 72.9+j42.4 & 71.4+j24.3  \\ 67.1 +j7.6  & 71.4+j24.3 & 72.9+j42.4 \end{array} \right], \nonumber
\end{equation} 
and
\begin{equation}\label{eq:imp_matrix2}
\vec{Z}_\mathrm{r}(0.5\lambda)=\left[ \begin{array}{ccc} 72.9 +j42.4 &-12.5 -j29.8 & 4.0 +j17.7 \\ -12.5-j29.8 & 72.9 +j42.4 & -12.5 -j29.8  \\ 4.0+j17.7 & -12.5 -j29.8 & 72.9 +j42.4 \end{array} \right]. \nonumber
\end{equation} 
It can be observed that the severe mutual coupling in the case of $d=0.05\lambda$ results in off-diagonal elements whose values are closer to the diagonal elements than in the case of $d=0.5\lambda$, where the diagonal elements are more dominant. Despite this, the impact of coupling on capacity is not immediately obvious, since the impedance matrix is embedded in (\ref{eq:G_coupled}), and is conditioned by the load matrix ${\vec{Z}}_\mathrm{l}$. Therefore, we next provide numerical simulations to give more insight into the impact of mutual coupling on MIMO performance.

\begin{figure}
         \centering
                 \includegraphics[width=0.7\columnwidth]{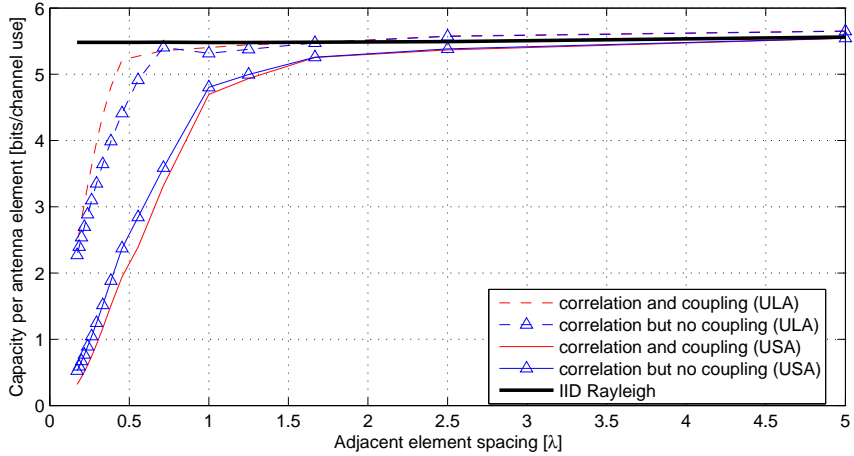}
         \caption{Impact of correlation and coupling on capacity per
                 antenna over different adjacent antenna spacing
                 for autonomous transmitters. $M=K$ and the
                 apertures of ULA and USA are $5\lambda$ and
                 $5\lambda \times 5\lambda$,
                 respectively.}
         \label{fig:coupling}
 \end{figure}

In MU-MIMO systems\footnote{We
remind the reader that in MU-MIMO systems, we replace $n_t$ and $n_r$
with $K$ and $M$ respectively.}, the terminals are autonomous so that we can assume that
the transmit array is uncoupled and uncorrelated. If the Kronecker model \cite{KeSP02} is assumed for the propagation channel
${\vec{G}}=\vec{\Psi}_\mathrm{r}^{1/2} \vec{G}_\mathrm{IID} \vec{\Psi}_\mathrm{t}^{1/2}$, 
where $\vec{\Psi}_\mathrm{t}$ and $\vec{\Psi}_\mathrm{r}$ are the transmit and receive 
correlation matrices, respectively, and $\vec{G}_\mathrm{IID}$ is the matrix with 
IID Rayleigh entries~\cite{FeFL08}. In this case, ${\vec{\Psi}}_\mathrm{t}^{1/2}=\vec{I}_K$ and $\vec{Z}_\mathrm{t}$ is diagonal.  For the particular case of
$M=K$, Figure \ref{fig:coupling} shows a plot of the uplink ergodic capacity (or average rate) per user, $C_\mathrm{mc}/K$, 
versus the antenna separation for ULAs with a fixed aperture of $5\lambda$ at the base station (with up
to $M=K=30$ elements). The correlation but no coupling case refers to the MIMO channel 
${\vec{G}}=\vec{\Psi}_\mathrm{r}^{1/2} \vec{G}_\mathrm{IID} \vec{\Psi}_\mathrm{t}^{1/2}$, 
whereas the correlation and coupling case refers to the effective channel matrix  ${\hat{\vec{G}}}_\mathrm{mc}$ in (\ref{eq:G_coupled}).
The environment is assumed to be uniform 2D Angular Power Spectrum (APS) and the
SNR is $\rho=20$ dB. The total power is fixed and equally 
divided among all users. 1000 independent realizations of the channel are used
to obtain the average capacity. For comparison, the corresponding ergodic capacity per user is also calculated for $K^2$ users 
and an $M^2$-element receive Uniform Square Array (USA) with $M=K$ and an
aperture size of $5\lambda \times 5\lambda$, for up to $M^2=900$
elements\footnote{Rather than advocating the practicality of $900$ users in a single cell, this assumption is only intended to demonstrate the limitation of aperture-constrained very large MIMO arrays at the base station to support parallel MU-MIMO channels.}. 

As can be seen in Figure \ref{fig:coupling}, the capacity per user begins to fall when the element spacing is reduced to below $2.5\lambda$ for the USAs, as opposed to below
$0.5\lambda$ for the ULAs, which shows that for a given antenna spacing, packing more elements in
more than one dimension results in significant degradation in capacity performance. Another distinction between the ULAs and USAs is that coupling is in fact beneficial for the capacity performance of ULAs with moderate antenna spacing (i.e. between $0.15\lambda$ and $0.7\lambda$), whereas for USAs the capacity with coupling is consistently lower than that with only correlation. The observed phenomenon for ULAs is similar to the behavior of two dipoles with decreasing element spacing~\cite{Lau10}. There, coupling induces a larger difference between the antenna patterns (i.e., angle diversity) over this range of antenna spacing, which helps to reduce correlation. At even smaller antenna spacings, the angle diversity diminishes and correlation increases. Together with loss of power due to coupling and impedance mismatch, the increasing correlation results in the capacity of the correlation and coupling case falling below that of the correlation only case, with the crossover occuring at approximately $0.15\lambda$. On the other hand, each element in the USAs experiences more severe coupling than that in the ULAs for the same adjacent antenna spacing, which inherently limits angle diversity.  

Even though Figure \ref{fig:coupling} demonstrates that both coupling and
correlation are detrimental to the capacity performance of very large
MIMO arrays relative to the IID case, it does not provide any specific information on the behavior of
${\hat{\vec{G}}}_{\mathrm{mc}}$. In particular, it is important to examine
the impact of correlation and coupling on the asymptotic orthogonality assumption made in (\ref{eq:G_mu_orthog})  for a very large array with a fixed aperture  in a MU setting. To this end, we
assume that the base station serves $K=15$ single antenna
terminals. The channel is normalized so that \textit{each} user terminal
has a reference SNR $\rho/K = 10$ dB in the SISO case with conjugate-matched single antennas. As before, the coupling and correlation at the base station is the result of implementing the antenna elements as a square array of fixed dimensions $5\lambda \times 5\lambda$ in a channel with uniform 2D APS.  The number of elements in the receive USA $M$ varies from 16 to 900, in order to support one dedicated channel per user. 

\begin{figure}
         \centering
                 \includegraphics[width=0.7\columnwidth]{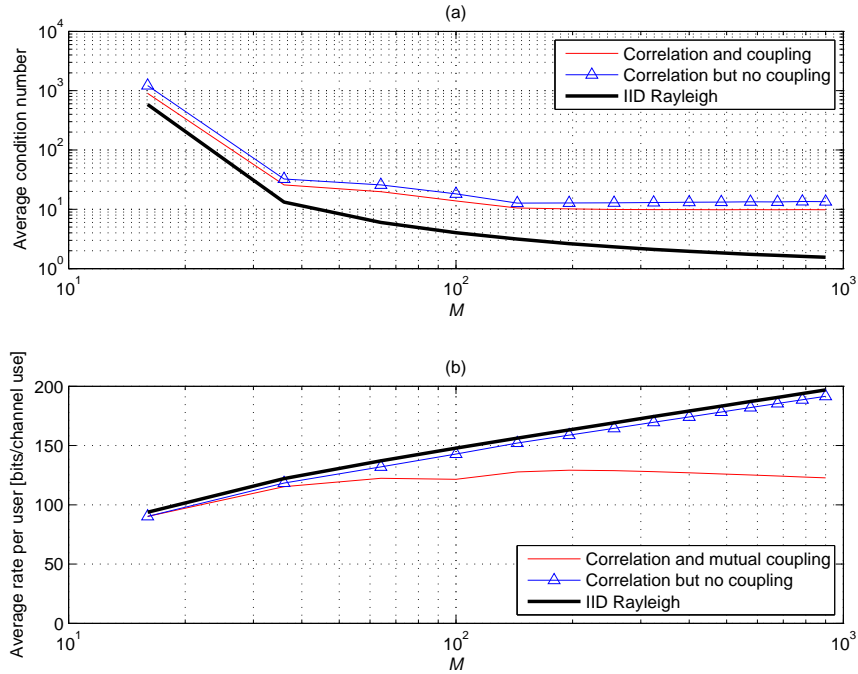}
         \caption{Impact of correlation and coupling on (a) asymptotic
                 orthogonality of the channel matrix and (b) max
                 sum-rate of the reverse link, for $K=15$.}
         \label{fig:asymptotic_error}
 \end{figure}

The average condition number of ${\hat{\vec{G}}_\mathrm{mc}\HT \hat{\vec{G}}_\mathrm{mc}}/K$ is given in Figure \ref{fig:asymptotic_error}(a) for 1000 channel
realizations. Since the propagation channel is assumed to be IID in (\ref{eq:G_coupled}) for simplicity, $\vec{D}_{\mathbf{\beta}}=\vec{I}_{K}$. This implies that the condition number of 
${\hat{\vec{G}}_\mathrm{mc}\HT \hat{\vec{G}}_\mathrm{mc}}/K$ should ideally approach one, which is observed for the IID Rayleigh case.
By way of contrast, it can be seen that the channel is not asymptotically orthogonal as assumed in (\ref{eq:G_mu_orthog}) in the presence of coupling and
correlation. The corresponding maximum rate for the reverse link per user is given in Figure \ref{fig:asymptotic_error}(b). It can be seen that if coupling is ignored, spatial correlation yields only a minor penalty, relative to the IID case. This is so because the transmit array of dimensions $5\lambda \times 5\lambda$ is large enough to offer almost the same number of spatial degrees of freedom ($K=15$)  as in the IID case, despite the channel not being asymptotically orthogonal. On the other hand, for the realistic case with coupling and correlation, adding more receive elements into the USA will eventually result in a reduction of  the achievable rate, despite having a lower average condition number than in the correlation but no coupling case. This is attributed to the significant power loss through coupling and impedance mismatch, which is not modeled in the correlation only case.

\subsection{Real propagation - measured channels} \label{sec:realpropagation}

When it comes to propagation aspects of MIMO as well as very large
MIMO the correlation properties are of paramount interest, since those
together with the number  of antennas at the terminals and base
station determines the orthogonality of the propagation channel matrix
and the possibility to separate different users or data streams. In
conventional MU-MIMO systems the ratio of number of base station antennas
and antennas at the terminals is usually close to 1, at least it rarely
exceeds 2. In very large MU-MIMO systems this ratio may very well exceed
100; if we also consider the number of expected simultaneous users,
$K$, the ratio at least usually exceeds 10. This is important because
it means that we have the potential to achieve a very large spatial
diversity gain. It also means that the distance between the
null-spaces of the different users is usually  large and, as mentioned
before, that the singular values of the tall propagation matrix tend
to have stable and large values. This is also true in the case where
we consider multiple users where we can consider each user as a part
of a larger distributed, but un-coordinated, MIMO system. In such a
system each new user ``consumes'' a part of the available diversity. Under certain reasonable assumptions and favorable propagation conditions, it will, however, still be possible to create a full rank
propagation channel matrix (\ref{eq:G_mu_orthog}) where all the eigenvalues have large
magnitudes and show a stable behavior. The question is now what we mean by the
statement that the propagation conditions should be favorable? One
thing is for sure: As compared to a conventional MIMO system, the
requirements on the channel matrix to get good performance in very large MIMO are relaxed to a large extent due to the tall structure of the matrix. 

It is well known in conventional MIMO modeling that scatterers tend to appear in groups with similar delays, angle-of-arrivals and angle-of-departures and they form so-called clusters. Usually the number of active clusters and distinct scatterers are reported to be limited, see e.g. \cite{Correia06}, also when the number of physical objects is large. The contributions from individual multipath components belonging to the same cluster are often correlated which reduces the number of effective scatterers. Similarly it has been shown that a cluster seen by different users, so called joint clusters, introduces correlation between users also when they are widely separated \cite{Poutanen10}. It is still an open question whether the use of large arrays makes it possible to resolve clusters completely, but the large spatial resolution will make it possible to split up clusters in many cases. There are measurements showing that a cluster can be seen differently from different parts of a large array \cite{Santos10}, which is beneficial since the correlation between individual contributions from a cluster then is decreased. 

\begin{figure}
        \centering
        \psfrag{IID 6x128}{\scalebox{.7}{ IID $6\!\!\times\!\! 128$ }}
        \psfrag{IID 6x6}{\scalebox{.7}{ IID $6\!\!\times\!\! 6$ }}
        \psfrag{meas 6x128}{\scalebox{.7}{ Meas $6\!\!\times\!\! 128$}}
        \psfrag{meas 6x6}{\scalebox{.7}{ Meas $6\!\!\times\!\! 6$}}
        \psfrag{ordered singular values [dB]}{\scalebox{.75}{\hspace*{-3mm} Ordered
        eigenvalues of $\vec{G}\HT \vec{G}$ [dB]}}
                \includegraphics[width=0.7\columnwidth]{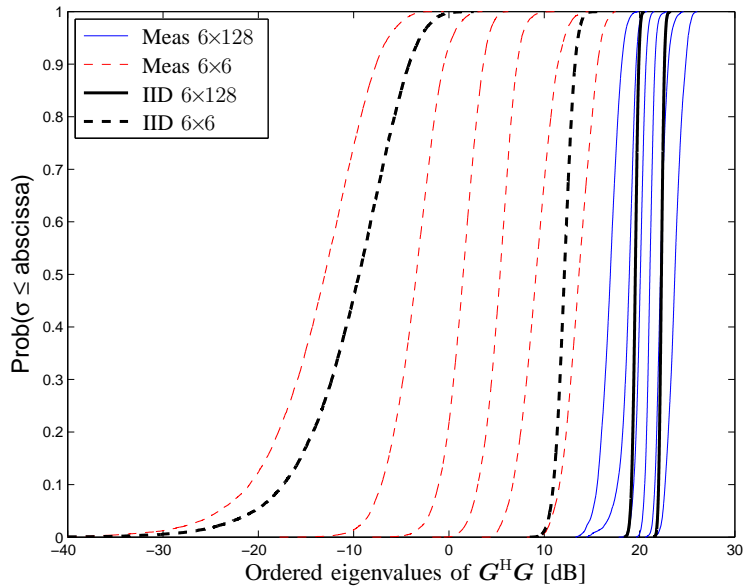}
        \caption{CDFs of ordered eigenvalues for a measured $6
                \times 128$ large array system, a measured $6 \times
                6$ MIMO system and  simulated IID $6 \times 6$ and
                $6\times 128$ MIMO systems. Note that for the simulated IID cases, 
                only the CDFs of the largest and smallest eigenvalues are shown for clarity. }
        \label{fig:SingValCDF}
\end{figure}

To exemplify the channel properties in a real situation we consider a
measured channel matrix where we have an indoor 128-antenna base
station consisting of four stacked double polarized 16 element
circular patch arrays, and 6 single antenna users. Three of the users are
indoors at various positions in an adjacent room and 3 users are
outdoors but close to the base station. The measurements were
performed at 2.6 GHz with a bandwidth of 50 MHz. In total we consider
an ensemble of 100 snapshots (taken from a continuous movement of the
user antenna along a 5-10 m line) and 161 frequency points, giving us
in total 16100 narrow-band realizations. It should be noted, though,
that they are not fully independent due to the non-zero coherence
bandwidth and coherence distance. The channels are normalized to
remove large scale fading and to maintain the small scale fading. The
mean power over all frequency points and base station antenna elements
is unity for all users. In Figure \ref{fig:SingValCDF} we plot the
Cumulative Distribution Functions (CDF) of the ordered
eigenvalues of $\vec{G}\HT \vec{G}$ (the leftmost solid curve
corresponds to the CDF of 
the smallest eigenvalue etc.)
for the $6 \times 128$ propagation matrix (``Meas 6x128''), together with
the corresponding CDFs for a $6 \times 6$ measured conventional MIMO
(``Meas 6x6'') system (where we have used a subset of 6 adjacent
co-polarized antennas on the base station). As a reference we also
plot the distribution of the largest and smallest eigenvalues for a simulated $6 \times 128$ and  $6 \times 6$ conventional MIMO system
(``IID 6x128'' and ``IID 6x6'') with independent identically distributed
complex Gaussian entries. Note that, for clarity of the figure, the
eigenvalues are not normalized with the number of antennas at the
base station and therefore there is an offset of
$10\log_{10}(M)$. This offset can be interpreted as a beamforming gain.
In any case, the relative spread of the eigenvalues is of more interest than their absolute levels. 

It can be clearly  seen that the large array provides eigenvalues
that all show a stable behavior (low variances) and have a relatively
low spread (small distances between the CDF curves). The difference
between the smallest and largest eigenvalue is only around
7 dB, which could be compared with the  conventional $6 \times 6$ MIMO
system where this difference is around 26 dB. 
This eigenvalue
spread corresponds to that of a 6x24 conventional MIMO system with IID
complex Gaussian channel matrix entries. Keeping in mind the
circular structure of the base station antenna array and that half of the elements are cross polarized, this
number of 'effective' channels is about what one could anticipate to get.
One important factor in realistic channels, especially for the uplink,
is that the received power levels from different users are not
equal. Power variations will increase both the eigenvalue spread and
the variance, and will result in a matrix that still is approximately
orthogonal, but where the diagonal elements of $\vec{G}\HT \vec{G}$
have varying mean levels, namely the $\vec{D}_{\beta}$ matrix in (\ref{eq:G_mu_orthog}).
 
\section{Transceivers}
We next turn our attention to the design of practical transceivers. A
method to acquire CSI at the base station
begins the discussion. Then we discuss precoders and detection
algorithms suitable for very large MIMO arrays.

\subsection{Acquiring CSI at the base station}
In order to do multiuser precoding in the forward link and detection
in the reverse link, the base station
must acquire CSI. Let us assume that the frequency response of the channel is
constant over $N_{\mathrm{Coh}}$ consecutive subcarriers. With small antenna arrays, one
possible system design is to let the
base station antennas transmit pilot symbols to the receiving
units. The receiving units perform channel estimation and feed back,
partial or complete, CSI via dedicated feedback channels. Such a
strategy does not rely on channel reciprocity (i.e., the forward channel should be the transpose of the reverse channel).  However, with a limited coherence
time, this strategy is not viable for large arrays. The number of time slots
devoted to pilot symbols must be at least as large as the number of
antenna elements at the base station divided by $N_{\mathrm{Coh}}$. When $M$ grows, the time spent
on transmitting pilots may surpass the coherence time of the channel.

Consequently, large antenna array technology must rely on channel reciprocity.
With channel reciprocity, the receiving units send pilot symbols via
TDD. Since the frequency response is assumed constant over
$N_{\mathrm{Coh}}$ subcarriers, $N_{\mathrm{Coh}}$ terminals can
transmit pilot symbols simultaneously during 1 OFDM symbol
interval. In total, this requires $K/N_{\mathrm{Coh}}$ time
slots (we remind the reader that $K$ is the number of terminals served). The  base station in the $k$-th cell
constructs its channel estimate ${\hat{\vec{G}}}\transp_{kk}$,
subsequently used for precoding in the forward link, based on the pilot
observations.  The power of each
pilot symbol is denoted $\rho\pil$.

\subsection{Precoding in the forward link: Collection of results for single cell systems}\label{precoderssinglecell}
User $k$ receives the $k$-th component of the
composite vector 
$$\vec{x}\fw=\vec{G}\transp\vec{s}\fw+\vec{w}\fw.$$
The vector $\vec{s}\fw$ is a precoded version of the data symbols
$\vec{q}\fw$. Each component of $\vec{s}\fw$ has average power
$\rho\fw/M$. Further, we assume that the channel matrix
$\vec{G}$ has
IID $\mathcal{CN}(0,1)$ entries. In what follows, we derive 
SNR/SINR (Signal-to-Interference-plus-Noise-Ratio) expressions for a
number of popular precoding techniques in the large system limit,
i.e., with
$M,K\to \infty$, but with a fixed ratio $\alpha=M/K$. The obtained expressions are tabulated in Table \ref{snrtable}.

Let us first discuss the performance of an Interference Free (IF) system
which will  subsequently  serve as a benchmark reference. The
best  performance that can be imagined will result if all
the channel energy to terminal $k$ is delivered to terminal $k$ without any
inter-user interference. In that case,  terminal $k$ receives the sample $x\fwk$ 
$$x\fwk = \sqrt{\sum_{\ell=1}^{M}|g_{\ell k}|^2}\,q\fwk+w\fwk.$$ 
Since
$\left(\sum_{\ell=1}^{M}|g_{\ell k}|^2\right)/M\to 1,\;M\to\infty,$ and
$\mathbb{E}\left\{q\fwk q\fwk\HT\right\}=\rho\fw/K,$
 the SNR per receiving unit for
IF systems converges to $\rho\fw \alpha$ as $M\to\infty$.


We now move on to practical precoding methods. The conceptually simplest
 approach is to invert the channel by means of the pseudo-inverse. This is
referred to as Zero-Forcing (ZF) precoding \cite{Peeletal1}. A variant of zero forcing is Block
Diagonalization \cite{ChoiMurch04}, which is not covered in this
paper. Intuitively, when  $M$ grows, $\vec{G}$ tends to have nearly orthogonal columns as the terminals are not
correlated due to their physical separation. This assures that the
performance of ZF precoding will be close to that of the IF
system. However, a disadvantage of ZF is that processing cannot be
 done distributedly at each antenna separately. With ZF precoding, all
 data must  instead be collected at a central node that handles the processing.

Formally, the ZF precoder sets
$$\vec{s}\fw=\frac{1}{\sqrt{\gamma}}(\vec{G}\transp)^+\vec{q}\fw=\frac{1}{\sqrt{\gamma}}\vec{G}^{\ast}(\vec{G}\transp
\vec{G}^{\ast})^{-1}\vec{q}\fw,$$
where the superscript ``+'' denotes the pseudo-inverse of a matrix,
i.e. $(\vec{G}\transp)^+=\vec{G}^\ast(\vec{G}\transp \vec{G}^\ast)^{-1}$, and $\gamma$
normalizes the average power in $\vec{s}\fw$ to $\rho\fw$.  A suitable choice for $\gamma$ is
$\gamma=\mathrm{Tr}(\vec{G}\transp \vec{G}^\ast)^{-1}/K$ which averages 
fluctuations in transmit power due to $\vec{G}$ but not to
$\vec{q}\fw$. The received sample $x\fwk$
with ZF precoding becomes
$$x\fwk = {\frac{q\fwk}{\sqrt{\gamma} }}+w\fwk.$$ 
With that, the instantaneous received SNR per terminal equals
\bea \label{ZF} \mathrm{SNR}&=&\frac{\rho\fw}{K\,\gamma} \nonumber \\
&=&\frac{\rho\fw}{\mathrm{Tr}(\vec{G}\transp \vec{G}^\ast)^{-1}}.
\eea

When both the number of terminals $K$ and the number of base station
antennas $M$ grow large, but with fixed  ratio $\alpha=M/K$,
$\mathrm{Tr}(\vec{G}\transp \vec{G}^\ast)^{-1}$ converges to a fixed
  deterministic value \cite{HWAllerton02}
\be \label{tracelimit} \mathrm{Tr}(\vec{G}\transp \vec{G}^\ast)^{-1}\to \frac{1}{\alpha -1},\quad \mathrm{as}\;
K,\, M\to \infty, \quad \frac{M}{K}=\alpha.\ee
Substituting (\ref{tracelimit}) into (\ref{ZF}) gives the expression in
Table \ref{snrtable}.
The conclusion is that ZF precoding achieves an $\mathrm{SNR}$ that
tends to the optimal $\mathrm{SNR}$ for an IF system with
$M-K$ transmit antennas when
the array size grows. Note that when $M=K$, one gets $\mathrm{SNR}=0$.

A problem with ZF precoding is that the construction of the pseudo-inverse
$(\vec{G}\transp)^+=\vec{G}^\ast(\vec{G}\transp \vec{G}^\ast)^{-1}$ requires
the inversion of a $K\times K$ matrix, which is
computationally expensive.  However, as $M$
grows,  $(\vec{G}\transp \vec{G}^\ast)/M$ tends to the identity
matrix, which has a trivial inverse. Consequently, the ZF precoder
tends to $\vec{G}^\ast$, which is nothing but a MF. This suggests that matrix inversion may not be needed when the
array is scaled up, as the MF precoder approximates the ZF precoder
well. Formally, the MF sets 
$$\vec{s}\fw=\frac{1}{\sqrt{\gamma}}\vec{G}^\ast\vec{q}\fw,$$
with $\gamma=\mathrm{Tr}(\vec{G}\transp \vec{G}^\ast)/K$.
A few simple manipulations lead to an asymptotic expression of the SINR,
which is given in Table \ref{snrtable}.

From the MF precoding SINR expression, it is
seen that the SINR can be made as high as
desired by scaling up the antenna array. However,  the MF precoder
exhibits an error floor since as $\rho\fw \to
\infty,\;\mathrm{SINR}\to \alpha$. 

We next turn the attention to scenarios where the base station has
imperfect CSI. Let $\hat{\vec{G}}\transp$ denote the Minimum Mean Square
Error (MMSE) channel estimate of the
forward link. The estimate satisfies,
$$\hat{\vec{G}}\transp=\xi \vec{G}\transp+\sqrt{1-\xi^2}\vec{E},$$
where $0\leq \xi\leq 1$ represents the reliability of the estimate and
$\vec{E}$ is a matrix with IID $\mathcal{CN}(0,1)$ distributed
entries.  SINR expressions for MF and ZF precoding are given in Table \ref{snrtable}.
 For any reliability $\xi$, the
$\mathrm{SINR}$ can be made as high as desired by scaling
up the antenna array. 



\begin{table}[th]
\begin{minipage}[c]{\linewidth}
\renewcommand{\thefootnote}{\thempffootnote}
\setcounter{footnote}{0}
\begin{center}
\begin{tabular}{|c||c|c|} 
 \multicolumn{3}{c}{SNR and SINR expressions as $K,M\to
  \infty,\quad M/K=\alpha$}\\ \hline \hline
Precoding Technique&Perfect CSI&Imperfect CSI\\
\hline \hline
Benchmark: IF System& $\rho\fw \alpha$&\\ \hline
Zero Forcing&$\rho\fw (\alpha-1)$& $\frac{\xi^2\,\rho\fw(\alpha-1)}{(1-\xi^2)\,\rho\fw+1}$\\\hline

Matched Filter & $\frac{\rho\fw \alpha}{\rho\fw +1}$& $\frac{\xi^2\rho\fw \alpha}{\rho\fw +1}$\\\hline

Vector Perturbation& $\approx
\frac{\rho\fw\,\alpha\,\pi}{6}\left(1-\frac{1}{\alpha}\right)^{1-\alpha},\quad
\alpha\lessapprox1.79$&N.A.\\
\hline\hline
\end{tabular}
\end{center}
\caption{SNR and SINR expressions for a collection of standard
  precoding techniques. }
\label{snrtable}\end{minipage}
\end{table}

Non-linear precoding techniques, such as DPC, Vector Perturbation (VP)
\cite{Peeletal05}, and lattice-aided methods  \cite{WindpassingerHuber}
are important techniques when $M$ is not much larger than $K$. This is
true since in
the $M\approx K$ regime, the performance gap of ZF to the IF benchmark is
significant, see Table \ref{snrtable}, and there is room for
improvement by non-linear techniques. However, the gap of ZF to an IF
system scales as $\alpha/(\alpha-1)$. When $M$ is, say, two times $K$,
this gap is only 3 dB. Non-linear techniques will operate closer to the
IF benchmark, but cannot surpass it. Therefore the gain of non-linear
methods does not at all justify the complexity increase.
The measured $6 \times 128$ channels that we discussed earlier in the paper behave as if
$\alpha\approx 4$. Hence, linear precoding is virtually optimal and
one can dispense with DPC.

For completeness we give an approximate large limit SNR
expression for VP, derived from the results of  \cite{RyanVP}, in
Table \ref{snrtable}. The expression is strictly speaking an upper bound
to the SNR, but is  reasonably tight \cite{RyanVP} so that it can be taken as an
approximation. For $\alpha\gtrapprox 1.79$, the SINR expression
surpasses that of an IF system, which makes the expression
meaningless. However, for larger values of $\alpha$, linear precoding
performs well and there is not much gain in using VP anyway.
For VP,  no SINR expression is
available in the literature with imperfect CSI.

\begin{figure}[t]
\begin{center}
  \scalebox{.7}{\includegraphics{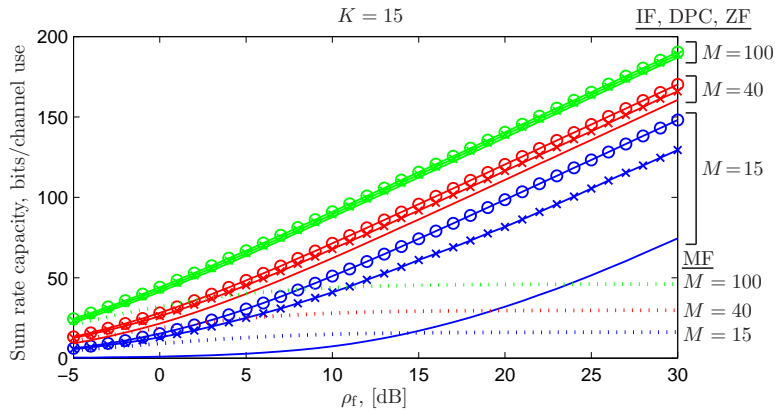}}
\vspace*{-3mm}  
\caption{\label{rates} Sum-rate capacities of single cell
  multiuser MIMO precoding techniques. The channel is IID complex
  Gaussian $\mathcal{CN}(0,1)$, there are $K=15$ terminals. Circles
  show the performance of IF systems, x-es refer to DPC, solid
  lines refer to ZF, and the dotted lines refer to MF.}
\end{center}
\vspace*{-2mm}  
\end{figure}

In Figure \ref{rates} we show ergodic sum-rate capacities for MF
precoding, ZF precoding, and DPC. As  benchmark performance we also show the
ensuing sum-rate capacity from an IF system. In
all cases, $K=15$ users are served and we show results for $M=15,\,
40,\,100$. For $M=15$, it can be seen that DPC decisively outperforms ZF
and is about 3 dB away from the IF benchmark
performance.  But
as $M$ grows, the advantage of DPC quickly diminishes. With $M=40$,
the gain of DPC is about 1 dB. This confirms that the performance gain does not at
all justify the complexity increase. With 100 base station antennas, ZF
precoding performs almost as good as an interference free system.
At low SNR, MF precoding is better than ZF precoding. It is
interesting to observe that this is true over a wide range of SNRs for
the case of $M=K$. Sum-rate capacity expressions of VP are currently not
available in the literature, since the optimal distribution of the
inputs for VP is not known to date.

\subsection{Precoding in the forward link: The ultimate limit of non-cooperative multi cell MIMO
  with large arrays}\label{downlinkultlimit}
\begin{figure}[t]
 \begin{center}
 \psfrag{Gm}{\scalebox{1}{\hspace*{-0mm} SIR [dB]}}
 \scalebox{.9}{\includegraphics{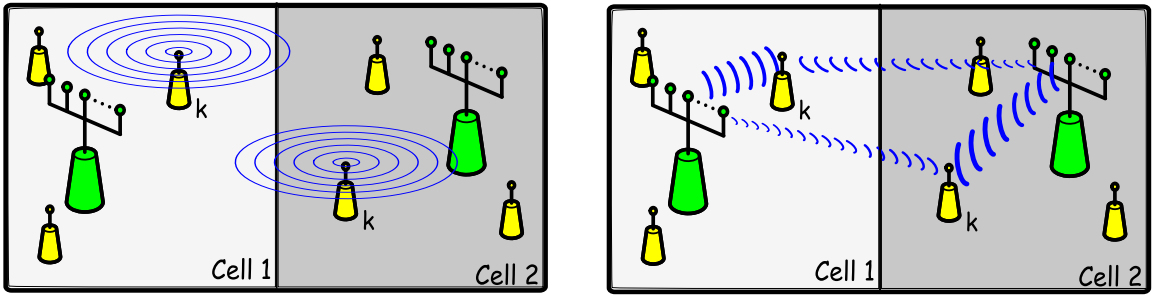}}
 \vspace*{-3mm}  
 \caption{\label{Pilotcont}  Illustration of the pilot contamination 
   concept. Left: During the training phase, the base station
   in cell 1 overhears the pilot transmission from other
   cells. Right: As a consequence, the transmitted vector from
   base station 1 will be partially \emph{beamformed} to the terminals in cell 2. }
 \end{center}
 \vspace*{-2mm}  
 \end{figure}

In this section, we investigate the limit of non-cooperative cellular
multiuser MIMO systems as $M$ grows without limit. The presentation
summarizes and extends the results of \cite{Marzetta10}. For single cell as well as for multi
cell MIMO, the end effect of letting $M$ grow without limits is that thermal
noise and small scale Rayleigh fading vanishes. However, as we will
discuss in detail, with multiple
cells the
interference from other cells due to pilot contamination does not
vanish. The concept of pilot contamination is novel in a cellular
MU-MIMO context and is illustrated in
Figure \ref{Pilotcont}, but was an issue in the context of CDMA, usually under the name
``pilot pollution''. The channel estimate computed by the
base station in  cell 1 gets contamined
from the pilot transmission of cell 2. The base station in cell 1 will
in effect beamform its signal partially along the
channel to the terminals in cell 2. Due to the beamforming, the
interference to cell 2 does not vanish asymptotically as
$M\to\infty$.

\begin{figure}[t]
 \begin{center}
 \psfrag{Gm}{\scalebox{1}{\hspace*{-0mm} SIR [dB]}}
 \scalebox{1}{\includegraphics{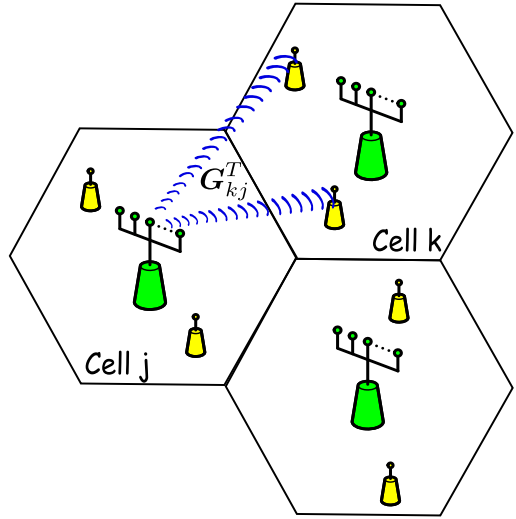}}
 \vspace*{-3mm}  
 \caption{\label{Cellstr} The composite channel between the base station
 in cell $j$ and the terminals in cell $k$ is denoted $\vec{G}_{k j}\transp$.}
 \end{center}
 \vspace*{-7mm}  
 \end{figure}

We consider a cellular  multiuser MIMO-OFDM system with hexagonal
cells and $N_{\mathrm{FFT}}$ subcarriers. All cells serves $K$
autonomous terminals
and has $M$ antennas at the base station. 
Further, a sparse scenario $K\leq M$ is assumed for simplicity. Hence, terminal scheduling aspects are not
 considered. The base stations are assumed non-cooperative. The
 $M\times K$ composite channel matrix between the $K$ terminals in cell $k$ and the
 base station in cell $j$ is denoted $\vec{G}_{k j}$. Relying on
 reciprocity, the forward link channel matrix between the base station
 in cell $j$ and the terminals in cell $k$ becomes $\vec{G}_{k
   j}\transp$ 
 (see Figure \ref{Cellstr}).

The base station in the $k$-th cell transmits the vector $\vec{s}\fwk$
which is a precoded version of the data symbols $\vec{q}\fwk$ intended
for the terminals in cell $k$. Each terminal in the $k$-th cell receives his
respective component of the composite vector
\be\label{chmodmumimo}\vec{x}\fwk=\rho\fw\sum_{j}\vec{G}\transp_{kj}\vec{s}_{\mathrm{f}j}+\vec{w}\fwk.\ee

As before, each element of  $\vec{G}_{k j}$ comprises a
small scale Rayleigh fading factor as well as a large scale factor that accounts
for geometric attenuation and  shadow fading. With that,
$\vec{G}_{k j}$ factors as
\be \label{propmodel}\vec{G}_{kj}=\vec{H}_{kj}\vec{D}_{\beta_{kj}}^{1/2}.\ee
In (\ref{propmodel}), $\vec{H}_{kj}$ is a $M\times K$ matrix which represents the small scale
fading between the terminals in cell $k$ to the base station in cell $j$,
all entries are IID $\mathcal{CN}(0,1)$ distributed. The $K\times K$ matrix
$\vec{D}_{\beta_{kj}}^{1/2}$ is a diagonal matrix comprising the elements
$\vec{\beta}_{kj}=[\beta_{kj1},\,\beta_{kj2},\ldots,\,\beta_{kjK}]$
along its main diagonal; each value $\beta_{kj\ell}$ represents the
large scale fading between terminal $\ell$ in
the $k$-th cell and the base station in cell $j$. 

The base station in the $n$-th cell processes its pilot observations
and obtains a channel estimate ${\hat{\vec{G}}}\transp_{nn}$ of
$\vec{G}\transp_{nn}$. In the worst case, the pilot signals
 in all other cells are perfectly synchronized with the pilot signals in
cell $n$. Hence, the channel estimate ${\hat{\vec{G}}}\transp_{nn}$ gets
contamined from pilot signals in other cells,
\be
\label{chest}{{\hat{{\vec{G}}}}\transp}_{nn}=\sqrt{\rho\pil}\vec{G}_{nn}\transp
+\sqrt{\rho\pil}\sum_{i\neq n}{\vec{G}}\transp_{in}+\vec{V}_{n}\transp.\ee
In (\ref{chest}) it is implicitly assumed that all terminals transmits
identical pilot signals. Adopting different pilot signals in different cells
does not improve the situation much \cite{Marzetta10} since the pilot
signals must at least be confined to the same signal space, which is
of finite dimensionality.

Note that, due to the geometry of the cells, $\vec{G}_{nn}$ 
is generally stronger than $\vec{G}_{in},\,i\neq n$. $\vec{V}_{n}$ is a matrix of
receiver noise during the training phase, uncorrelated with all
propagation matrices, and comprises  IID $\mathcal{CN}(0,1)$ distributed elements;
$\rho\pil$ is a measure of the SNR during of the pilot transmission phase.

Motivated by the virtual optimality of simple linear precoding from Section 
\ref{precoderssinglecell}, we let the base station in cell $n$ use the MF
$(\hat{\vec{G}}\transp_{nn})\HT={\hat{{\vec{G}}}}_{nn}^\ast$ as precoder. We later investigate  zero-forcing
precoding. Power normalization of the precoding matrix is unimportant when
$M\to\infty$ as will become clear shortly. The $\ell$-th terminal in the $j$-th cell
receives the $\ell$-th component of the vector
$\vec{x}_{\mathrm{f}j}=[x_{\mathrm{f}j1},\,x_{\mathrm{f}j1},\ldots,\,x_{\mathrm{f}jK}]\transp$.
Inserting (\ref{chest}) into (\ref{chmodmumimo}) gives 
\bea \label{recsig}\vec{x}_{\mathrm{f}j}&=&
\sqrt{\rho\fw}\sum_{n}\vec{G}\transp_{jn}{\hat{\vec{G}}}^\ast_{nn}\,\vec{q}_{\mathrm{f}n}+\vec{w}_{\mathrm{f}j}
\nonumber \\
&=&\sqrt{\rho\fw}\sum_{n}\vec{G}\transp_{jn}\left[\sqrt{\rho\pil}\sum_{i}{\vec{G}}\transp_{in}+\vec{V}_n\transp\right]\HT\,\vec{q}_{\mathrm{f}n}+\vec{w}_{\mathrm{f}j}.\eea
The composite received signal vector $\vec{x}_{\mathrm{f}j}$ in
(\ref{recsig}) contains terms of the form
$\vec{G}\transp_{jn}\vec{G}^\ast_{in}.$
As $M$ grows large, only terms  where $j=i$ remain significant. We
get
$$\frac{\vec{x}_{\mathrm{f}j}}{M\sqrt{{\rho\fw\rho\pil}}}\to\sum_{n}\frac{\vec{G}\transp_{jn}\vec{G}^\ast_{jn}}{M}\vec{q}_{\mathrm{f}n},\quad \mathrm{as}\;M\to\infty.$$
Further, as $M$ grows, the effect of small scale Rayleigh fading
vanishes,
$$\frac{\vec{G}_{jn}\transp \vec{G}_{jn}^\ast}{M}\to {\vec{D}}_{\beta_{jn}}.$$
 Hence, the processed
received signal of the $\ell$-th receiving unit in the $j$-th cell is
\be \label{recsig_unlimited} \frac{x_{\mathrm{f}j\ell}}{M\sqrt{\rho\fw\rho\pil}}\to \beta_{jj \ell}q_{\mathrm{f}j\ell}+\sum_{n\neq j}
\beta_{jn\ell }q_{\mathrm{f}n\ell}.\ee
The SIR of terminal $\ell$ becomes
\be \label{mfsir}\mathrm{SIR}=\frac{{\beta^2_{jj\ell}}}{\sum_{n\neq j}
 {\beta^2_{jn\ell }}},\ee
which does not contain any thermal noise or small scale fading
 effects! Note that devoting more power to the training phase does not decrease
 the pilot contamination effect and leads to the same SIR. This is a
 consequence of the worst-case-scenario assumption that the pilot transmissions in all
 cells overlap. If the pilot transmissions are staggered so that pilots
in one cell collide with data in other cells, devoting more power to
 the training phase is indeed beneficial. However, in a multi cell
 system, there will always be some pilot transmissions that collide, although
 perhaps not in neighboring cells.

We now replace the MF precoder in (\ref{recsig}) with the
pseudo-inverse of the channel estimate $({\hat{{\vec{G}}}_{nn}}\transp)^+=\hat{{\vec{G}}}_{nn}^\ast(\hat{{\vec{G}}}\transp_{nn}\hat{{\vec{G}}}_{nn}^\ast)^{-1}.$
Inserting the expression for the channel estimate (\ref{chest})
gives
$$({\hat{{\vec{G}}}}\transp_{nn})^+=\left[\sqrt{\rho\pil}\sum_{i}{\vec{G}}^\ast_{in}+\vec{V}^\ast_{n}\right]\left(\left[\sqrt{\rho\pil}\sum_{i^{\prime}}{\vec{G}}\transp_{i^{\prime}n}+\vec{V}\transp_{n}\right]\left[\sqrt{\rho\pil}\sum_{i^{\prime\prime}}{\vec{G}}^\ast_{i^{\prime\prime}n}+\vec{V}^\ast_{n}\right]\right)^{-1}.$$
Again, when $M$ grows, only products of correlated terms remain
significant,
$$({\hat{{\vec{G}}}}\transp_{nn})^+\to\frac{1}{M\rho\pil}\left[\sqrt{\rho\pil}\sum_{i}{\vec{G}}_{in}^\ast+\vec{V}^\ast_{n}\right]\left(\sum_i \vec{D}_{\beta_{in}}+\frac{1}{\rho\pil}\vec{I}_K\right)^{-1}.$$

The processed composite received vector in the $j$-th cell becomes
$$\sqrt{\frac{\rho\pil}{\rho\fw}}\vec{x}_{\mathrm{f}j} \to \sum_n
{\vec{D}}_{\beta_{jn}}\left(\sum_i \vec{D}_{\beta_{in}}+\frac{1}{\rho\pil}\vec{I}_K\right)^{-1}\vec{q}_{\mathrm{f}n}.  $$
Hence, the $\ell$-th receiving unit in the $j$-th cell receives
$$\sqrt{\frac{\rho\pil}{\rho\fw}}x_{\mathrm{f}j\ell}\to \frac{\beta_{jj\ell}}{\sum_{i}\beta_{ ij\ell}+\frac{1}{\rho\pil}}q_{\mathrm{f}j\ell}+\sum_{n\neq j}
\frac{\beta_{jn\ell}}{\sum_i \beta_{in\ell}+\frac{1}{\rho\pil}}q_{\mathrm{f}n\ell}.$$
The SIR of terminal $k$ becomes
\be \label{zfsir}\mathrm{SIR}=\frac{\beta_{jj\ell}^2\,/\,(\sum_{i}\beta_{ ij\ell}+\frac{1}{\rho\pil})^2}{\sum_{n\neq j}
\beta_{jn\ell}^2\,/\,(\sum_i \beta_{in\ell}+\frac{1}{\rho\pil})^2}.\ee
We point out that with ZF precoding, the ultimate limit is independent
        of $\rho\fw$ but not of $\rho\pil$. As
        $\rho\pil\to 0$, the performance of the ZF precoder converges
        to that of the MF precoder.

Another popular technique is to first regularize
 the matrix $\hat{{\vec{G}}}\transp_{nn}\hat{{\vec{G}}}_{nn}^\ast$ before
 inverting \cite{Peeletal1}, so that the precoder is given by
$$\hat{{\vec{G}}}_{nn}^\ast(\hat{{\vec{G}}}\transp_{nn}\hat{{\vec{G}}}_{nn}^\ast+\delta\,\vec{I}_K)^{-1},$$ 
 where $\delta$ is a parameter subject to optimization. Setting
 $\delta=0$ results in the ZF precoder while $\delta\to\infty$ gives the
 MF precoder. For single cell systems, $\delta$ can be chosen
 according to \cite{Peeletal1}. For multi cell
 MIMO, much less is known, and  we briefly
 elaborate on  the impact of $\delta$ with simulations that will be
 presented later. We point out that the effect of
 $\rho\pil$ can be removed by taking $\delta=-M/{\rho\pil}$.

The ultimate limit can be further improved by adopting a power
allocation strategy at the base stations. Observe that we only study
non-cooperative base stations. In a distributed MIMO system, i.e. the
processing for several base stations is carried out at a central
processing unit, ZF could be applied across the base stations to reduce the effects of the
pilot contamination. This would imply an estimation of the
  factors $\{\beta_{kj\ell}\}$, which is feasible since they are
  slowly changing and are assumed to be constant over
  frequency.

\subsubsection{Numerical results} \label{numresprec}
We assume that each base station serves $K=10$ terminals. The cell diameter
(to a vertex) is 1600 meters and no terminal is allowed to get closer
to the base station than
100 meters. The large scale fading factor
$\beta_{kj\ell}$ decomposes as
$\beta_{kj\ell }=z_{kj\ell}/r_{kj\ell}^{3.8},$
where $z_{kj \ell}$ represents the shadow fading and abides a log-normal distribution (i.e. $10\log_{10}(z_{kj\ell })$ is zero-mean Gaussian distributed with
standard deviation $\sigma_{\mathrm{shadow}}$) with
$\sigma_{\mathrm{shadow}}=8$ dB and $r_{kj\ell}$ is the distance
between the base station in the $j$-th cell and terminal $\ell$ in the
$k$-th cell. Further, we assume a frequency reuse factor of 1.

\begin{figure}[t]
\begin{center}
\psfrag{SINRdb}{\scalebox{1}{\hspace*{-2mm} SIR [dB]}}
\psfrag{cumulativedistribution}{\scalebox{1}{\hspace*{-6mm} Cumulative Distribution}}
\psfrag{ABCDEFGHIJKLMNOPQRS}{\scalebox{.83}{ZF precoder}}
\psfrag{data2}{\scalebox{.83}{MF precoder }}
\psfrag{data3}{\scalebox{.83}{Regularized ZF, $\delta\!=\!M/20$}}
\scalebox{.7}{\includegraphics{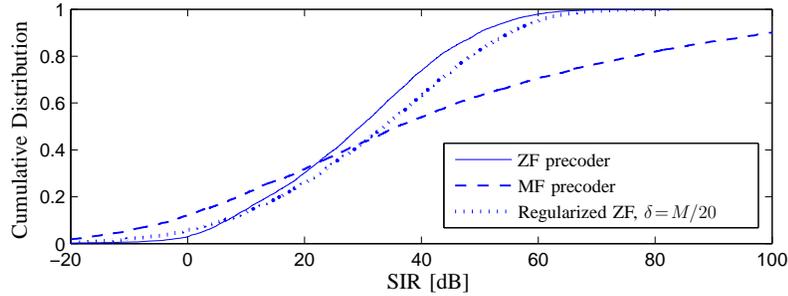}}
\vspace*{-3mm}  
\caption{\label{fi1} Cumulative distributions on the SIR for the MF
  precoder, the ZF precoder, and a regularized ZF precoder with
  $\delta=M/20$. The number of terminals served is $K=10$.}
\end{center}
\vspace*{-7mm}  
\end{figure}

Figure \ref{fi1} shows CDFs
of the SIR as $M$ grows without limit. We plot the SIR for MF
precoder (\ref{mfsir}), the ZF precoder (\ref{zfsir}), and a
regularized ZF precoder with $\delta=M/20$.
From the figure, we see that the distribution of the SIR is more
concentrated around its mean for ZF precoding compared with MF
precoding. However, the mean capacity
$\mathbb{E}\{\log_2(1+\mathrm{SIR})\}$
is larger for the MF precoder than for the ZF precoder (around 13.3 bits/channel use compared to
9.6 bits/channel use). With a regularized ZF precoder, the mean
capacity and outage probability are traded against eachother.

\begin{figure}[t]
\begin{center}
\psfrag{SNRinDB}{\scalebox{1}{\hspace*{-4mm} SIR [dB]}}
\psfrag{M}{\scalebox{1}{\hspace*{0mm}$M$}}
\psfrag{ABCDEFGHIJKLMNO}{\scalebox{.83}{MF.
    $\mathbb{E}\{\mathrm{SIR}\}$, $M\to\infty$}}
\psfrag{data2}{\scalebox{.83}{ZF. $\mathbb{E}\{\mathrm{SIR}\}$, $M\to\infty$}}
\psfrag{data4}{\scalebox{.83}{ZF. $\mathbb{E}\{\mathrm{SIR}\}$}}
\psfrag{data3}{\scalebox{.83}{MF. $\mathbb{E}\{\mathrm{SIR}\}$}}
\scalebox{.7}{\includegraphics{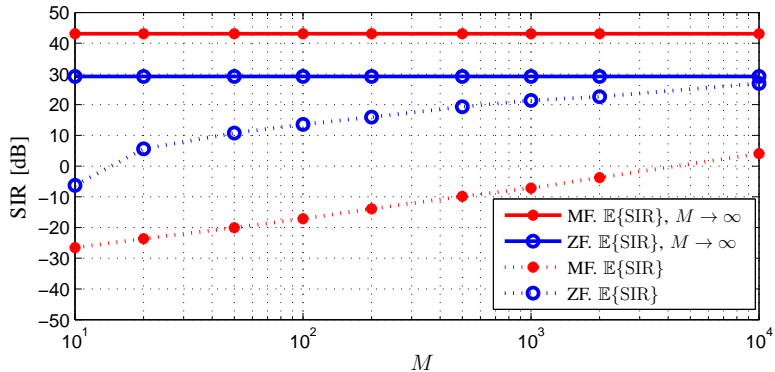}}
 \vspace*{-3mm}  
\caption{\label{fi2} Signal-to-interference-ratios for MF and ZF
  precoders as a function of $M$. The two uppermost curves are
  asymptotic mean values of the SIR as $M\to\infty$. The bottom two 
  curves show mean values of the SIR for finite $M$. The number of
  terminals served is $K=10$. }
 \end{center}
 \vspace*{-7mm}  
\end{figure}

We next consider finite values of $M$. In Figure \ref{fi2} 
the SIR for MF and ZF precoding is plotted against $M$ for infinite
 SNRs $\rho\pil$ and $\rho\fw$. With
'infinite' we mean that the SNRs are large enough so that the
performance is limited by pilot contamination.
The two uppermost curves show the mean SIR as $M\to\infty$. As can
be seen, the limit is around 11 dB higher with MF precoding. The two
bottom curves show the mean SIR for MF and ZF precoding for finite
$M$. The ZF precoder decisively outperforms the MF
precoder and achieves a hefty share of the asymptotic limit with
around 10-20 base station antenna elements per terminal. In order to reach
a given mean SIR, MF precoding requires at least two orders of
magnitude more base station antenna elements than ZF precoding does. 

In the particular case $\rho\pil=\rho\fw=10$ dB, the SIR of the MF precoder is
about 5 dB worse compared with infinite $\rho\pil$ and $\rho\fw$
over the entire range of $M$ showed in Figure \ref{fi2}. Note that as  $M\to\infty$, this loss will vanish.

\subsection{Detection in the reverse link: Survey of algorithms for single cell systems}
\label{sec:uplink_detection}
Similarly to in the case of MU-MIMO precoders,  
simple linear detectors are close to optimal if $M\gg K$ under
 favorable propagation conditions.  
However, operating points with $M\approx K$ are also important in practical systems with many users. 
Two more advanced categories of methods, iterative filtering schemes and random step methods,
have recently been proposed for detection in the very large MIMO regime.
We compare these methods with  the linear methods and to tree search methods in the following.
The fundamentals of the schemes are explained
for hard-output detection, experimental results are provided, and 
soft detection is discussed at the end of the section.
Rough computational complexity estimates for the presented methods are given in Table \ref{tab:complexity}.

\subsubsection{Iterative linear filtering schemes}
\label{sec:mat_inv}
These methods work by resolving the detection of the signaling vector $\vec{q}$ by iterative linear filtering, and   at each iteration by means of new propagated information
from the previous estimate of $\vec{q}$. The propagated information can be either hard, i.e., consist of decisions on the signal vectors,
or soft, i.e., contain some probabilistic measures of the transmitted symbols (observe that here, soft information is propagated between
different iterations of the hard detector). The methods typically 
employ matrix inversions repeatedly during the iterations, which, if the inversions occur frequently, may be computationally heavy  when $M$ is large. 
Luckily, 
the matrix inversion lemma can be used to remove some of the complexity stemming from matrix inversions.

As an example of a soft information-based method, we describe the
conditional MMSE with soft interference cancellation
(MMSE-SIC) scheme \cite{lampe:99}.  
 The algorithm is
initialized with a linear MMSE estimate $\tilde{\vec{q}}$ of $\vec{q}$. Then
for each user $k$, an interference-canceled signal $\vec{x}_{i,k}$,
where subscript $i$ is the iteration number, is
  constructed by removing inter-user interference. Since the estimated
  symbols at each iteration are not perfect, there will still be
  interference from other users in the signals $\vec{x}_{i,k}$. 
This interference is modeled as Gaussian and the residual interference plus noise power is estimated. Using this estimate, an MMSE filter 
conditioned on filtered output from the previous iteration
is computed
  for each user $k$. The bias is removed and a soft MMSE estimate of
  each symbol given the filtered output, is propagated to the next iteration. The algorithm iterates these steps a
  predefined number $N_{\rm{Iter}}$ of times.

Matrix inversions need to be computed for every 
realization $\vec{x}$, every user symbol $q_k$,  and every
iteration. Hence the number of matrix inversions per decoded vector is $KN_{\rm{Iter}}$.
One can employ the matrix inversion lemma in order to reduce the number of matrix inversions to 1 per iteration.
The idea is to formulate the inversion for user $k$ as a rank one
update of a general inverse matrix at each iteration.

The BI-GDFE algorithm \cite{Liang:06} is
equation-wise similar to MMSE-SIC
\cite{Liang:08}.
Compared to MMSE-SIC, it has 
two differences. The linear MMSE filters of MMSE-SIC depend on the received vector $\vec{x}$, while the 
BI-GDFE filters, which are functions of a parameter that varies with iteration, the so-called input-decision correlation (IDC), do not.
This means that for a channel $\vec{G}$ that is fixed for many signaling vectors, all filters, which still vary for the different users and iterations, can be 
precomputed. Further, BI-GDFE propagates hard instead of soft decisions.

\subsubsection{Random step methods}
\label{sec:inv_free}
The methods categorized in this section are matrix-inversion-free, except possibly for the initialization stage, where 
the MMSE solution is usually used.
A basic matrix inversion-free search method starts with the initial vector, and evaluates the MSE for vectors in its 
 neighborhood with $N_{\text{Neigh}}$ vectors. 
 The neighboring vector with smallest MSE is
 chosen, and the process restarts, and continues like this for $N_{\text{Iter}}$ iterations. 
The Likelihood Ascent Search (LAS) algorithm \cite{vardhan:08} only permits transitions to states with 
lower MSE, and converges monotonically to a local minima in this way. 
An upper bound of bit error rate and a lower bound on asymptotic multiuser efficiency for the LAS detector were presented in \cite{sun:las}.

Tabu Search (TS) \cite{Zhao:07} is superior to the
LAS algorithm in that it permits transitions to states with larger MSE 
values, and it can in this way avoid local minima. TS also keeps a list 
 of recently traversed signaling vectors, with maximum number of entries $N_{\rm{Tabu}}$, that are temporarily forbidden moves, 
as a means for moving away to new areas of the search space.
 This strategy gave rise to the algorithm's name.

\subsubsection{Tree-based algorithms}
\label{sec:fcsd}
The most prominent algorithm within this class is the Sphere Decoder
(SD) \cite{FP,Larsson2009}. The SD is in fact an ML decoder, 
but which only considers points inside a sphere with certain radius. If the sphere
is too small for finding any signaling points, it has to be increased.
Many tree-based low-complexity algorithms try to reduce the search by
only expanding the fraction of the tree-nodes that appear the most ``promising''. 
One such method is the stack decoder \cite{Salah:08}, where the nodes of
the tree are expanded in the order of least Euclidean distance to the
received signal.
The average complexity of the sphere decoder is however exponential in
$K$ \cite{jalden}, and SD is thus not suitable in the large MIMO
regime where  $K$ is large.

The Fixed Complexity Sphere Decoder (FCSD) \cite{FCSD}  is a low-complexity,
suboptimal, version of the SD. 
All combinations of the first, say $r$, scalar symbols in $\vec{q}$ are enumerated,
 i.e., with a full search, and for each such combination, 
the remaining $K-r$
symbols are detected by means of ZF-DF.
This implies that the FCSD is highly
parallelizable since $|\mathcal{S}|^r$ hardware chains can be used,
 and further, it has a constant complexity.
 A sorting algorithm employing the matrix inversion lemma for finding which symbols  should be processed
 with full complexity and which ones should be detected with ZF-DF can
 be found in \cite{FCSD}.

The FCSD eliminates columns from the matrix
$\vec{G}$, which implies that the matrix gets better conditioned,
which in turn boosts the performance of linear
detectors. For $M\gg K$, the channel
matrix is, however,  already well conditioned, so the situation does not improve
much by eliminating a few columns. Therefore, the FCSD should mainly be used
in the case of $M\approx K$.

\begin{table}[th]
\begin{minipage}[c]{\linewidth}
\begin{center}
\begin{tabular}{|c||c||c|} 
\hline \hline 
Detection technique & Complexity for each realization of $\vec{x}$ &Complexity for each realization of $\vec{G}$\\
\hline \hline
MMSE         &$MK$                &    $MK^2+K^3$  \\ \hline
MMSE-SIC     &$(M^2K+M^3)N_{\text{Iter}}$              &                                                          \\ \hline
BI-GDFE      &$MKN_{\text{Iter}}$ & $(M^2K+M^3)N_{\text{Iter}}$           \\ \hline 
TS           &$((M+N_{\text{Tabu}})N_{\text{Neigh}}+MK)N_{\text{Iter}}$ &$MK^2+K^3$                                              \\ \hline
FCSD         &$(M^2+K^2+r^2)|\mathcal{S}|^r$                    &        $MK^2+K^3$                          \\ \hline
MAP          &$MK|\mathcal{S}|^K$&                                                              \\ \hline
\end{tabular}
\end{center}
\caption{Rough complexity estimates for detectors in terms of floating point operations. 
If a significant amount of the computations in question can be pre-processed for each $\vec{G}$ in slow fading,
the pre-processing complexity is given in the right column.
}
\label{tab:complexity}
\end{minipage}
\vspace*{-8mm}
\end{table}

\subsubsection{Numerical comparisons of the algorithms}
\label{sec:detector_experiments}
We now compare the detection algorithms described above experimentally. 
QPSK is used in all simulations and Rayleigh fading is assumed, i.e., the channel matrix is chosen to have
independent components which are distributed as $\mathcal{CN}(0,1)$. 
The transmit power is denoted $\rho$.
In all experiments, simulations are run until 500 symbol errors are counted. 
We also add an interference-free (IF) genie solution, that enjoys the same receive signaling power as
the other methods, without multi-user interference.

As mentioned earlier, when there is a large excess of base station 
antennas, simple linear detection performs well. It is natural to ask
for the number $\alpha=M/K$ when this effect kicks in. To give a
feel for this, we show the uncoded BER performance versus
$\alpha$, for the particular case of $K=15$, in Figure
\ref{fig:GMM2}. 
For the measurements in Figure \ref{fig:GMM2}, we let $\rho \sim 1/M$. 
MMSE-SIC uses $N_{\text{Iter}}=6$, BI-GDFE uses $N_{\text{Iter}}=4$ since further iterations gave no improvement,
and the IDC parameter was chosen from preliminary simulations.
The TS neighborhood is
  defined as the closest modulation points \cite{Zhao:07}, 
and TS uses $N_{\text{Iter}}\!=\!N_{\text{Tabu}}\!=\!60$.
For FCSD, we choose $r=8$.
We observe that when the ratio $\alpha$ is above 5 or so, 
the simple linear MMSE method performs well, while there is room for
improvements by more advanced detectors when $\alpha<5$.

Since we saw in Figure \ref{fig:GMM2} that there is a wide range of $\alpha$ where MMSE is
largely sub-optimal, we now consider the case $M=K$.
Figure \ref{fig:det_complexity} shows 
comparisons of uncoded BER of the studied detectors as functions of their complexities (given in Table \ref{tab:complexity}).
We consider the case without possibility of pre-processing, i.e., the column entries in Table \ref{tab:complexity} are summed for each scheme,
$M=K=40$, and we use $\rho=12$ dB.
We find that TS and MMSE-SIC perform best. For example, at a BER of 0.002, the TS is
1000 times less complex than the FCSD.

Figure \ref{fig:detector_snr} shows a plot of BER versus transmit signaling power $\rho$ for
$M=K=40$, when the scheme parameters are the maximum values in the experiment in Figure \ref{fig:det_complexity}.
It is seen that TS and MMSE-SIC perform best across the
entire SNR range presented. Note that the  ML
detector, with a search space of size $2^{80}$, cannot outperform the
IF benchmark. Hence, remarkably, we can conclude that TS and MMSE-SIC are
operating not more than 0.9 dB away from the ML
detector for $40\times 40$ MIMO.

\subsubsection{Soft-input soft-output detection}
\label{sec:detector_experiments-soft}
\begin{figure}
\begin{center}
\psfrag{TS}[l]{\tiny TS}
\psfrag{MMSE-SIC}[l]{\tiny MMSE-SIC}
\psfrag{BI-GDFE}[l]{\tiny BI-GDFE} 
\psfrag{genie}[l]{\tiny IF}
\psfrag{MMSE}[l]{\tiny MMSE}
\psfrag{FCSD}[l]{\tiny FCSD}
\psfrag{BER}[cc][cc]{\scriptsize BER}
\psfrag{alpha}[cc][cc]{\scriptsize $\alpha$}
  {\includegraphics[width=7.8cm]{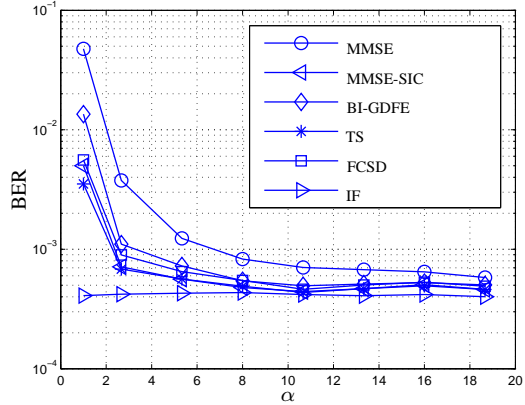}}\hfil
 \vspace*{-3mm}  
\caption{Comparisons of BER for $K=15$ and varying values of $\alpha$. }
\label{fig:GMM2}
\end{center}
\vspace*{-6mm}
\end{figure}

\begin{figure}
\begin{center}
\psfrag{TS}[l]{\tiny TS}
\psfrag{MMSE-SICaaaaa}[l]{\tiny MMSE-SIC}
\psfrag{BI-GDFE}[l]{\tiny BI-GDFE}
\psfrag{genie}[l]{\tiny IF}
\psfrag{MMSE}[l]{\tiny MMSE}
\psfrag{FCSD}[l]{\tiny FCSD}
\psfrag{BER}[cc][cc]{\scriptsize BER}
\psfrag{Number of floating point operations}[cc][cc]{\scriptsize Number of floating point operations}
   {\includegraphics[width=7.8cm]{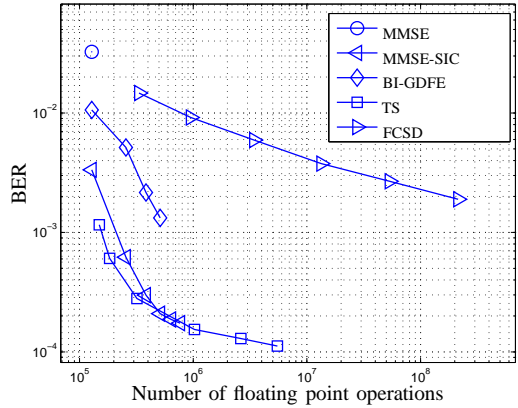}}\hfil
 \vspace*{-3mm}  
\caption{Comparisons of BER of the studied detectors  as functions of their complexities given in Table \ref{tab:complexity}.
We consider the case without possibility of pre-processing, i.e., the column entries in Table \ref{tab:complexity} are summed for each scheme.
The number of antennas $M=K=40$, and transmit signaling power $\rho=12$ dB.}
\label{fig:det_complexity}
\end{center}
\vspace*{-6mm}
\end{figure}

\begin{figure}
\begin{center}
\psfrag{TS}[l]{\tiny TS}
\psfrag{MMSE-SICaaaaa}[l]{\tiny MMSE-SIC}
\psfrag{BI-GDFE}[l]{\tiny BI-GDFE}
\psfrag{genie}[l]{\tiny IF}
\psfrag{MMSE}[l]{\tiny MMSE}
\psfrag{FCSD}[l]{\tiny FCSD}
\psfrag{BER}[cc][cc]{\tiny BER}
\psfrag{Transmit SNR (dB)}[cc][cc]{\scriptsize  $\rho$ [dB]}
\includegraphics[width=7.8cm]{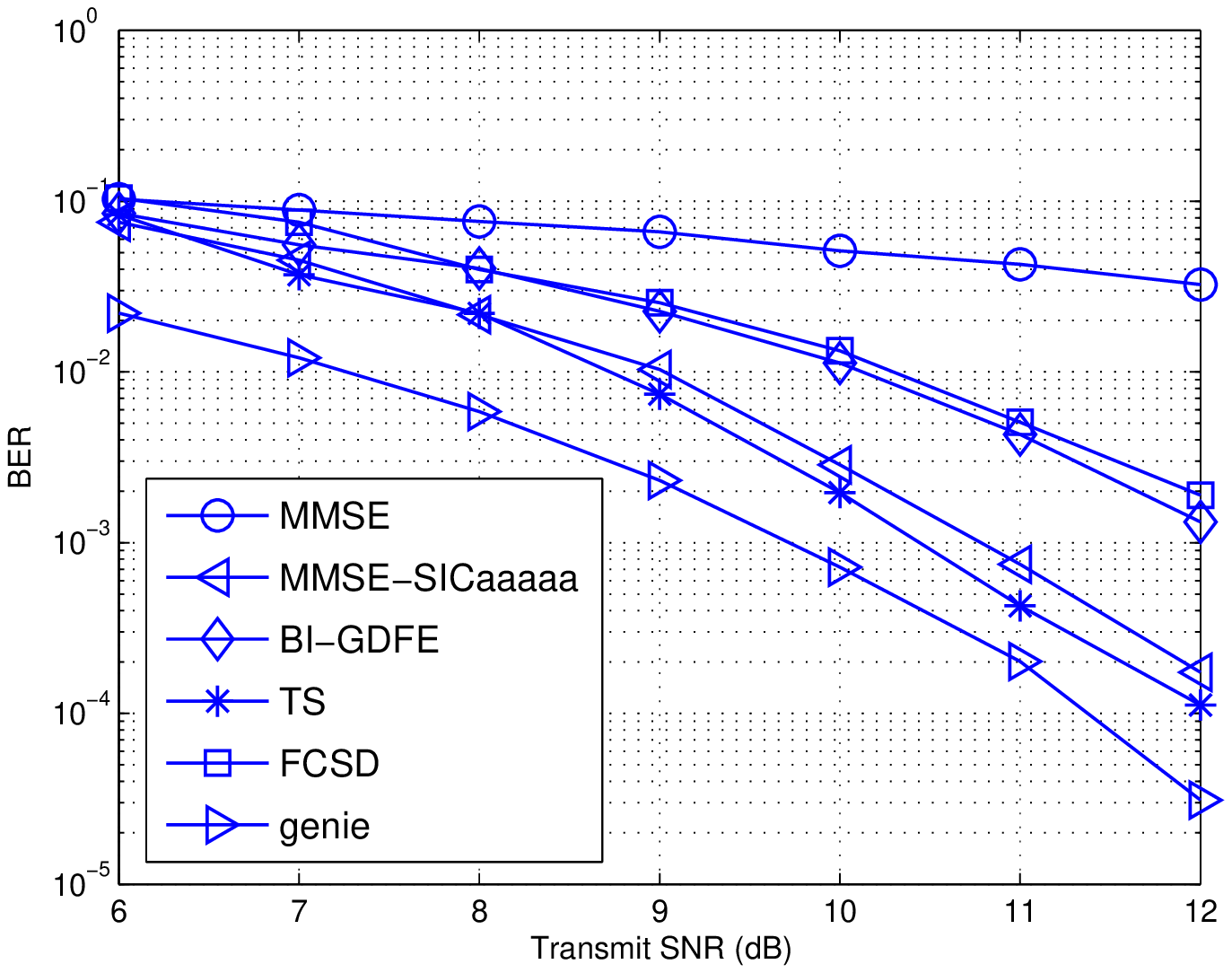}
 \vspace*{-3mm}  
\caption{Comparisons of the of the studied detectors for different transmit signaling power $\rho$. 
The scheme parameters are the maximum values in Figure
\ref{fig:det_complexity} and the number of antennas is $M=K=40$. }
\label{fig:detector_snr}
\end{center}
\vspace*{-6mm}
\end{figure}

The hard detection schemes above are easily evolved to soft detection methods. 
One should not in general draw conclusions about soft detection from hard detection.
Literature investigating schemes similar to the ones above, but operating in the coded large system limit,
are in agreement with Figures \ref{fig:GMM2}, \ref{fig:det_complexity}, and \ref{fig:detector_snr}.
In \cite{Verdu:99}, analytic CDMA spectral efficiency expressions for both MF, ZF, and linear MMSE, are given. 
The results are the following.  
In the limit of large ratios $\alpha$, all three methods perform likewise, and as well as the optimum joint detector and
CDMA with orthogonal spreading codes.
For $\alpha \approx 20$, MF starts to perform much worse than the other methods.
At $\alpha \approx 4/3$, ZF performs drastically worse than MMSE, but the MMSE method loses
significantly in performance compared to joint processing. 

With MMSE-SIC, a-priori information is easily incorporated in the MMSE filter derivation by conditioning.
This requires the computation of the filters for each user,
each symbol interval, and each decoder iteration \cite{Caire2004}. Another MMSE filter is derived by
unconditional incorporation of the a-priori probabilities, which results in MMSE filters
 varying for each user and iteration, similarly to for BI-GDFE above.  
Density evolution analysis of conditional and unconditional MMSE-SIC in a CDMA setting, and in the limit of infinite $N$ and $K$, shows that their
coded BER waterfall region can occur within two dB  from that of the MAP detector \cite{Caire2004}.  
In terms of spectral efficiency, the MAP detector and conditional and unconditional MMSE-SIC perform likewise.

For random step and tree-based methods, the main problem is to obtain a good list of candidate $\vec{q}$-vectors for approximate LLR evaluation, where
all bits should take the values 0 and 1 at least once. 
With the TS and FCSD methods, we start from lists containing the hard detection results and the vectors searched to achieve this result,
for creating an approximate max-log LLR. If a bit value for a bit position is missing, or if higher accuracy is needed, 
one can add vectors in the vicinity of the obtained set, see \cite{FCSDmap}.
A soft-output version of the LAS algorithm has been shown to operate
around 7 dB away from capacity in a coded 
V-BLAST setting with $M=K=600$  \cite{vardhan:08}.
Instead of using the max-log approximations for approximating LLR as in \cite{FCSDmap}, the PM algorithm keeps a sum of terms \cite{Persson:11}. 
There are many other approaches which may be suitable for
soft-output large scale MIMO detection, e.g.,  Markov chain Monte-Carlo techniques \cite{Zhu_MCMC:05}.

\section{Summary}
Very large MIMO offers the unique prospect within wireless
communication of saving an order of magnitude, or more, in transmit
power. As an extra bonus, the effect of small scale fading averages
out so that only the much more slowly changing large scale fading remains.
Hence, very large MIMO has the potential to bring radical changes to
the field. 

As the number of base station antennas grows, the system
gets almost entirely limited from the reuse of pilots in neighboring
cells, the so called \emph{pilot contamination} concept. This effect
appears to be a fundamental challenge of very large MIMO
system design, which warrants future research on the topic. 

We have also seen that the interaction between antenna elements can 
incur significant losses, both to channel orthogonality and link capacity. 
For large MIMO systems this is especially problematic since with a fixed 
overall aperture, the antenna spacing must be reduced. Moreover, the severity 
of coupling problem also depends on the chosen array geometry, e.g., linear 
array versus planar array. The numerical examples show that for practical 
antenna terminations (i.e., with no coupling cancellation), the primary impact 
of coupling is in power loss, in comparison to the case where only spatial 
correlation is accounted for. Notwithstanding, it is found that moderate 
coupling can help to reduce correlation and partially offset the impact of power 
loss on capacity. 

We have also surveyed uplink detection algorithms for cases where the
number of single antenna  users  and the number of base
station antennas is about the same, but both numbers are large, e.g. 40. The uplink detection problem
becomes extremely challenging in this case since the search space is exponential in
the number of users. By receiver tests and comparisons of several
state-of-the-art detectors, we have demonstrated that even this scenario
can be handled. 
Two especially promising detectors are
the MMSE-SIC and the TS, which both can operate very close to
the optimal ML detector.

To corroborate the theoretical models and claims of the paper, we have
also set up a small measurement campaign using an indoor 128 antenna element base
station and 6 single antenna users. In reality, channels are (generally) not IID, and thus there is a performance loss compared to ideal channels. However, the same trends appear and the measurements indicated a stable and robust performance. There are still many open issues with respect to the behavior in realistic channels that need further research and understanding, but the overall system performance seems very promising.

\subsection*{Sidebar: Approximate matrix inversion}\label{sec:approxmatrixinv}
Much of the computational complexity of the ZF-precoder and the reverse link
detectors lies in the
 inversion of a $K\times K$ matrix $\vec{Z}$. Although  base stations have high
 computational power, it is of interest to find approximate solutions by
 simpler means than outright  inversion.

In the following, we review an intuitive method for approximate matrix
inversion. It is known that
if a $K\!\!\times \!\!K$ matrix $\vec{Z}$ has the property
$$\lim_{n\to\infty}(\vec{I}_K-\vec{Z})^n=\vec{0}_K,$$
then its inverse can be expressed as a Neumann series \cite{Stewart98}
\be \label{neumann} \vec{Z}^{-1}=\sum_{n=0}^{\infty} (\vec{I}_K-\vec{Z})^n.\ee

Ostensibly, it appears that matrix inversion using (\ref{neumann}) is
even more complex than direct inversion since both matrix
inversion and  multiplication are $\mathcal{O}(K^3)$ operations.
However, in hardware, matrix multiplication is strongly preferred over inversion
 since it does not require any divisions. Moreover, if only the result
 of the inverse times a vector $\vec{s}=\vec{Z}^{-1}\vec{q}$ is
 of interest, then (\ref{neumann}) can be implemented as a series of
 cascaded matched filters. The complexity of each matched filter
 operation is only $\mathcal{O}(K^2)$.

Let us first consider the case of $K\times M$ matrix $\vec{G}$ with
independent and $\mathcal{CN}(0,1)$ distributed entries. We remind the
reader that $\alpha=M/K$. The objective
is now to approximate the inverse of the Wishhart matrix $\vec{Z}=\vec{G}\vec{G}\HT$.  As $K$ and $M$ grows, the
eigenvalues of $\vec{Z}$ converges to a fixed deterministic
distribution known as the Marchenko-Pastur distribution. The largest
and the smallest eigenvalues of $\vec{Z}$ converge to
$$\lambda_{\max}(\vec{Z})\to\left(1+\frac{1}{\sqrt{\alpha}}\right)^2,\quad\quad\lambda_{\min}(\vec{Z})\to\left(1-\frac{1}{\sqrt{\alpha}}\right)^2.$$
Some minor manipulations show that 
$$\lambda_{\max}\left(\frac{\alpha}{1+\alpha}\vec{Z}\right)\to1+2\frac{\sqrt{\alpha}}{1+\alpha},\quad\quad\lambda_{\min}\left(\frac{\alpha}{1+\alpha}\vec{Z}\right)\to1-2\frac{\sqrt{\alpha}}{1+\alpha}.$$
Hence, the eigenvalues of
$\vec{I}_K-\alpha/(1+\alpha)\vec{Z}=\vec{I}_K-\vec{Z}/(M+K)$ lie approximately in
the range $[-2\sqrt{\alpha}/(1+\alpha),2\sqrt{\alpha}/(1+\alpha)]$; note that
$2\sqrt{\alpha}/(1+\alpha)\leq 1$ whenever $\alpha>1$. Therefore
\be\label{matinv2}\lim_{n\to\infty}\left(\vec{I}_K-\frac{1}{M+K}\vec{Z}\right)^n=\vec{0}_K.\ee
When $M/K$ is large, say 5-10 or so,
(\ref{matinv2}) converges rapidly, and only a few terms needs to be computed.
For finite dimensions $K$ and $M$, the eigenvalues of a particular
channel realization can lie outside the range
$[-2\sqrt{\alpha}/(1+\alpha),2\sqrt{\alpha}/(1+\alpha)]$. Therefore an attenuation
factor $\delta<1$ is introduced.
 Altogether, the inverse of $\vec{G}=\vec{\vec{Z}}\vec{\vec{Z}}\HT$ can be
approximated as
\be
\label{matinv3}\vec{Z}^{-1}\approx\frac{\delta}{M+K}\sum_{n=0}^{L}\left(\vec{I}_K-\frac{\delta}{M+K}\vec{Z}\right)^n.
\ee

Replacing the weighting coefficent $1/(M+K)$ with
$c/\mathrm{Tr}(\vec{Z})$, $c$ a constant,  provides a robust method for matrix
approximation when the channel matrix has an unknown distribution. 
 Other techniques, e.g. based on the Cayley-Hamilton Theorem and random
matrix theory, have been
extensively used for CDMA receivers, see
\cite{VerduMuller01,Honig01}. If 
the weighting coefficients are optimized, the matrix inversion in
CDMA receivers can be approximated with only $\approx 8$ terms.

\bibliographystyle{IEEEtran}
\bibliography{bib/IEEEabrv,bib/refs}

\begin{thebibliography}{10}
\providecommand{\url}[1]{#1}
\csname url@rmstyle\endcsname
\providecommand{\newblock}{\relax}
\providecommand{\bibinfo}[2]{#2}
\providecommand\BIBentrySTDinterwordspacing{\spaceskip=0pt\relax}
\providecommand\BIBentryALTinterwordstretchfactor{4}
\providecommand\BIBentryALTinterwordspacing{\spaceskip=\fontdimen2\font plus
\BIBentryALTinterwordstretchfactor\fontdimen3\font minus
  \fontdimen4\font\relax}
\providecommand\BIBforeignlanguage[2]{{%
\expandafter\ifx\csname l@#1\endcsname\relax
\typeout{** WARNING: IEEEtran.bst: No hyphenation pattern has been}%
\typeout{** loaded for the language `#1'. Using the pattern for}%
\typeout{** the default language instead.}%
\else
\language=\csname l@#1\endcsname
\fi
#2}}

\bibitem{LTE}
E.~Dahlman, S.~Parkvall, J.~Sk{\"o}ld, and P.~Beming, \emph{3G Evolution HSPA
  and LTE for Mobile Broadband}.\hskip 1em plus 0.5em minus 0.4em\relax
  Academic Press, 2008.

\bibitem{tv:05}
D.~Tse and P.~Viswanath, \emph{Fundamentals of wireless communications}.\hskip
  1em plus 0.5em minus 0.4em\relax Cambridge University Press, 2005.

\bibitem{Larsson2009}
E.~G. Larsson, ``{MIMO detection methods: how they work},'' \emph{IEEE Signal
  Processing Magazine}, vol.~26, no.~3, pp. 91--95, 2009.

\bibitem{jalden}
J.~Jald{\'e}n and B.~Ottersten, ``On the complexity of sphere decoding in
  digital communications,'' \emph{IEEE Trans. Signal Processing}, vol.~53,
  no.~4, pp. 1474--1484, Apr. 2005.

\bibitem{Marzetta10}
T.~L. Marzetta, ``Noncooperative cellular wireless with unlimited numbers of
  base station antennas,'' \emph{IEEE Trans. Wireless. Commun.}, vol.~9,
  no.~11, pp. 3590--3600, Nov. 2010.

\bibitem{randommtrx}
A.~M. Tulino and S.~Verdu, \emph{Random Matrix Theory and Wireless
  Communications}.\hskip 1em plus 0.5em minus 0.4em\relax Now Publishers, 2004.

\bibitem{LRT2004}
G.~Lerosey, J.~de~Rosny, A.~Tourin, A.~Derode, G.~Montaldo, and M.~Fink, ``Time
  reversal of electromagnetic waves,'' \emph{Phys. Rev. Lett.}, vol.~92,
  no.~19, p. 193904, May 2004.

\bibitem{TWF1996}
J.-L. Thomas, F.~Wu, and M.~Fink, ``Time reversal focusing applied to
  lithotripsy,'' \emph{Ultrasonic Imaging}, vol.~18, no.~2, pp. 106--121, 1996.

\bibitem{Foschini1999}
G.~J. Foschini, ``Layered space-time architecture for wireless communication in
  a fading environment when using multi-element antennas,'' \emph{Bell Labs
  Technical Journal}, vol.~1, no.~2, 1999.

\bibitem{Mattaiou10}
M.~Matthaiou, M.~R. McKay, P.~J. Smith, and J.~A. Nossek, ``On the condition
  number distribution of complex {W}ishart matrices,'' \emph{IEEE Trans.
  Commun.}, vol.~58, no.~6, pp. 1705--1717, June 2010.

\bibitem{Jindal03}
S.~Vishwanath, N.~Jindal, and A.~Goldsmith, ``Duality, achievable rates, and
  sum-rate capacity of {MIMO} broadcast channels,'' \emph{IEEE Trans. Inform.
  Theory}, vol.~49, no.~10, pp. 2658--2668, Oct. 2003.

\bibitem{WSS06}
H.~Weingarten, Y.~Steinberg, and S.~Shamai, ``The capacity region of the
  {G}aussian multiple-input multiple-output broadcast channel,'' \emph{IEEE
  Trans. Inform. Theory}, vol.~52, no.~9, pp. 5011--5023, Sept. 2006.

\bibitem{HeB2004}
B.~E. Henty and D.~D. Stancil, ``Multipath-enabled super-resolution for {RF}
  and microwave communication using phase-conjugate arrays,'' \emph{Phys. Rev.
  Lett.}, vol.~93, no.~24, Dec 2004.

\bibitem{PoAK03}
T.~S. Pollock, T.~D. Abhayapala, and R.~A. Kennedy, ``Antenna saturation
  effects on {MIMO} capacity,'' in \emph{Proc. IEEE Int. Conf. Commun. (ICC)},
  vol.~4, May 2003, pp. 2301--2305.

\bibitem{WeGJ05}
S.~Wei, D.~Goeckel, and R.~Janaswamy, ``On the asymptotic capacity of {MIMO}
  systems with antenna arrays of fixed length,'' \emph{IEEE Trans. Wireless
  Commun.}, vol.~4, no.~4, pp. 1608--1621, July 2005.

\bibitem{HaFu06}
L.~Hanlen and M.~Fu, ``Wireless communication systems with-spatial diversity: a
  volumetric model,'' \emph{IEEE Trans. Wireless Commun.}, vol.~5, no.~1, pp.
  133--142, Jan. 2006.

\bibitem{Jana02}
R.~Janaswamy, ``Effect of element mutual coupling on the capacity of fixed
  length linear arrays,'' \emph{{IEEE} Antennas Wireless Propagat. Lett.},
  vol.~1, pp. 157--160, 2002.

\bibitem{Lau10}
B.~K. Lau, ``Multiple antenna terminals,'' in \emph{{MIMO}: From Theory to
  Implementation}, C.~Oestges, A.~Sibille, and A.~Zanella, Eds.\hskip 1em plus
  0.5em minus 0.4em\relax San Diego: Academic Press, 2011, pp. 267--298.

\bibitem{WaJe04a}
J.~W. Wallace and M.~A. Jensen, ``Mutual coupling in {MIMO} wireless systems: a
  rigorous network theory analysis,'' \emph{IEEE Trans. Wireless Commun.},
  vol.~3, no.~4, pp. 1317--1325, July 2004.

\bibitem{VoWS08}
C.~Volmer, J.~Weber, R.~Stephan, K.~Blau, and M.~A. Hein, ``An eigen-analysis
  of compact antenna arrays and its application to port decoupling,''
  \emph{{IEEE} Trans. Antennas Propagat.}, vol.~56, no.~2, pp. 360--370, Feb.
  2008.

\bibitem{LaAK06b}
B.~K. Lau, J.~B. Andersen, G.~Kristensson, and A.~F. Molisch, ``Impact of
  matching network on bandwidth of compact antenna arrays,'' \emph{{IEEE}
  Trans. Antennas Propagat.}, vol.~54, no.~11, pp. 3225--3238, Nov. 2006.

\bibitem{MoBB00}
A.~L. Moustakas, H.~U. Baranger, L.~Balents, A.~M. Sengupta, and S.~H. Simon,
  ``Communication through a diffusive medium: coherence and capacity,''
  \emph{Science}, vol. 287, pp. 287--290, Jan. 2000.

\bibitem{FeFL08}
Y.~Fei, Y.~Fan, B.~K. Lau, and J.~S. Thompson, ``Optimal single-port matching
  impedance for capacity maximization in compact {MIMO} arrays,'' \emph{{IEEE}
  Trans. Antennas Propagat.}, vol.~56, no.~11, pp. 3566--3575, Nov. 2008.

\bibitem{Bal05}
C.~A. Balanis, \emph{Antenna Theory - Analysis and Design}.\hskip 1em plus
  0.5em minus 0.4em\relax New Jersey: John Wiley \& Sons, 2005.

\bibitem{KeSP02}
J.~P. Kermoal, L.~Schumacher, K.~I. Pedersen, P.~E. Mogensen, and
  F.~Fredriksen, ``A stochastic {MIMO} radio channel model with experimental
  validation,'' \emph{IEEE J. Sel. Areas Commun.}, vol.~20, no.~6, pp.
  1211--1226, Aug. 2008.

\bibitem{Correia06}
L.~M. {Correia (ed)}, \emph{Mobile Broadband Multimedia Networks, Techniques,
  Models and Tools for 4G}.\hskip 1em plus 0.5em minus 0.4em\relax Academic
  Press, 2006.

\bibitem{Poutanen10}
J.~Poutanen, K.~Haneda, J.~Salmi, V.~M. Kolmonen, F.~Tufvesson, T.~Hult, and
  P.~Vainikainen, ``Significance of common scatterers in multi-link
  scenarios,'' in \emph{Proc. 4th European Conference on Antennas and
  Propagation (EuCAP 2010)}, Barcelona, Spain, Apr. 2010.

\bibitem{Santos10}
T.~Santos, J.~K\aa{}redal, P.~Almers, F.~Tufvesson, and A.~Molisch, ``Modeling
  the ultra-wideband outdoor channel - measurements and parameter extraction
  method,'' \emph{IEEE Trans. Wireless. Commun.}, vol.~9, no.~1, pp. 282--290,
  2010.

\bibitem{Peeletal1}
C.~B. Peel, B.~M. Hochwald, and A.~L. Swindlehurst, ``A vector-perturbation
  technique for near-capacity multiantenna communication --- part {I}: Channel
  inversion and regularization,'' \emph{IEEE Trans. Commun.}, vol.~53, no.~1,
  pp. 195--202, Jan. 2005.

\bibitem{ChoiMurch04}
L.-U. Choi and R.~D. Murch, ``A transmit preprocessing technique for multiuser
  {MIMO} systems using a decomposition approach,'' \emph{IEEE Trans. Wireless
  Commun.}, vol.~3, no.~1, pp. 20--24, Jan. 2004.

\bibitem{HWAllerton02}
B.~Hochwald and S.~Vishwanath, ``Space-time multiple access: linear growth in
  the sum rate,'' in \emph{Proc. 40th Annual Allerton Conf. Communications,
  Control and Computing}, Monticello, IL., Oct. 2002.

\bibitem{Peeletal05}
B.~M. Hochwald, C.~B. Peel, and A.~L. Swindlehurst, ``A vector-perturbation
  technique for near-capacity multiantenna communication --- part {II}:
  perturbation,'' \emph{IEEE Trans. Commun.}, vol.~53, no.~5, pp. 537--544, May
  2005.

\bibitem{WindpassingerHuber}
C.~Windpassinger, R.~F.~H. Fischer, and J.~B. Huber, ``Lattice-reduction-aided
  broadcast precoding,'' \emph{IEEE Trans. Commun.}, vol.~52, no.~12, pp.
  2057--2060, Dec. 2004.

\bibitem{RyanVP}
D.~J. Ryan, I.~B. Collings, I.~V.~L. Clarkson, and R.~W. Heath, ``Performance
  of vector perturbation multiuser {MIMO} systems with limited feedback,''
  \emph{IEEE Trans. Commun.}, vol.~57, no.~9, pp. 2633--2644, Sept. 2009.

\bibitem{lampe:99}
A.~Lampe and J.~Huber, ``On improved multiuser detection with iterated soft
  decision interference cancellation,'' in \emph{Proc. IEEE Communication
  Theory Mini-Conference}, June 1999, pp. 172--176.

\bibitem{Liang:06}
Y.-C. Liang, S.~Sun, and C.~K. Ho, ``Block-iterative generalized decision
  feedback equalizers for large {MIMO} systems: algorithm design and asymptotic
  performance analysis,'' \emph{IEEE Trans. Signal Processing}, vol.~54, no.~6,
  pp. 2035--2048, June 2006.

\bibitem{Liang:08}
Y.-C. Liang, E.~Y. Cheu, L.~Bai, and G.~Pan, ``On the relationship between
  {MMSE-SIC} and {BI-GDFE} receivers for large multiple-input multiple-output
  channels,'' \emph{IEEE Trans. Signal Processing}, vol.~56, no.~8, pp.
  3627--3637, Aug. 2008.

\bibitem{vardhan:08}
K.~Vishnu~Vardhan, S.~Mohammed, A.~Chockalingam, and B.~Sundar~Rajan, ``A
  low-complexity detector for large {MIMO} systems and multicarrier {CDMA}
  systems,'' \emph{IEEE J. Sel. Areas Commun.}, vol.~26, no.~3, pp. 473--485,
  Apr. 2008.

\bibitem{sun:las}
Y.~Sun, ``A family of likelihood ascent search multiuser detectors: an upper
  bound of bit error rate and a lower bound of asymptotic multiuser
  efficiency,'' \emph{IEEE J. Sel. Areas Commun.}, vol.~57, no.~6, pp.
  1743--1752, June 2009.

\bibitem{Zhao:07}
H.~Zhao, H.~Long, and W.~Wang, ``Tabu search detection for {MIMO} systems,'' in
  \emph{Proc. IEEE International Symposium On Personal, Indoor And Mobile Radio
  Communications (PIMRC)}, Sept. 2007, pp. 1--5.

\bibitem{FP}
U.~Fincke and M.~Pohst, ``Improved methods for calculating vectors of short
  length in a lattice, including a complexity analysis,'' \emph{Math. Comput.},
  vol.~44, pp. 463--471, Apr. 1985.

\bibitem{Salah:08}
A.~Salah, G.~Othman, R.~Ouertani, and S.~Guillouard, ``New soft stack decoder
  for {MIMO} channel,'' in \emph{Asilomar Conference on Signals, Systems and
  Computers}, Oct. 2008, pp. 1754--1758.

\bibitem{FCSD}
L.~Barbero and J.~Thompson, ``Fixing the complexity of the sphere decoder for
  {MIMO} detection,'' \emph{IEEE Trans. Wireless Commun.}, vol.~7, no.~6, pp.
  2131--2142, June 2008.

\bibitem{Verdu:99}
S.~Verdu and S.~Shamai, ``Spectral efficiency of {CDMA} with random
  spreading,'' \emph{IEEE Trans. on Information Theory}, vol.~45, no.~2, pp.
  622 --640, Mar. 1999.

\bibitem{Caire2004}
G.~Caire, R.~Muller, and T.~Tanaka, ``Iterative multiuser joint decoding:
  optimal power allocation and low-complexity implementation,'' \emph{IEEE
  Trans. Inform. Theory}, vol.~50, no.~9, pp. 1950--1973, Sept. 2004.

\bibitem{FCSDmap}
L.~Barbero and J.~Thompson, ``Extending a fixed-complexity sphere decoder to
  obtain likelihood information for turbo-{MIMO} systems,'' \emph{{IEEE} Trans.
  Veh. Technol.}, vol.~57, no.~5, pp. 2804--2814, Sep 2008.

\bibitem{Persson:11}
D.~Persson and E.~G. Larsson, ``Partial marginalization soft {MIMO} detection
  with higher order constellations,'' \emph{IEEE Transactions on Signal
  Processing}, vol.~59, no.~1, pp. 453--458, Jan. 2011.

\bibitem{Zhu_MCMC:05}
H.~Zhu, B.~Farhang-Boroujeny, and R.-R. Chen, ``On performance of sphere
  decoding and {M}arkov chain {M}onte {C}arlo detection methods,'' \emph{IEEE
  Signal Processing Letters}, vol.~12, no.~10, pp. 669--672, Oct. 2005.

\bibitem{Stewart98}
G.~Stewart, \emph{Matrix Algorithms: Basic decompositions.}\hskip 1em plus
  0.5em minus 0.4em\relax SIAM, 1998.

\bibitem{VerduMuller01}
R.~M{\"u}ller and S.~Verdu, ``Desing and analysis of low-compelxity
  interference mitigation on vector channels,'' \emph{IEEE J. Sel. Areas
  Commun.}, vol.~19, no.~8, pp. 1429--1441, Aug. 2001.

\bibitem{Honig01}
M.~L. Honig and W.~Xiao, ``Performance of reduced-rank linear interference
  suppression,'' \emph{IEEE Trans. Inform. Theory}, vol.~47, no.~5, pp.
  1928--1946, July 2001.

\end{thebibliography}
\end{document}